\pdfoutput=1
\documentclass[11pt,twoside,a4paper,cmspaper,final,collab]{cms-tdr}
\def\svnVersion{3b03036}\def\svnDate{2025/04/28}\def\cmsCernNoTag{CERN-EP-2024-293}\def\cmsCernDate{\today}\def\cmsMessage{Published in the Journal of High Energy Physics as \href{http://dx.doi.org/10.1007/JHEP04(2025)115}{\doi{10.1007/JHEP04(2025)115}.}}
\begin{document}\cmsNoteHeader{SMP-24-005}

\newcommand{\pp}{\ensuremath{\Pp\Pp}\xspace}
\renewcommand{\qqbar}{\ensuremath{\PQq\PAQq}\xspace}
\newcommand{\WZ}{\ensuremath{\PW\PZ}\xspace}
\newcommand{\ttZ}{\ensuremath{\ttbar\PZ}\xspace}
\newcommand{\ttX}{\ensuremath{\ttbar\PX}\xspace}
\newcommand{\ZZ}{\ensuremath{\PZ\PZ}\xspace}
\newcommand{\qqZZ}{\ensuremath{\PQq\PAQq\to\PZ\PZ}\xspace}
\newcommand{\ggZZ}{\ensuremath{\Pg\Pg\to\PZ\PZ}\xspace}
\newcommand{\WW}{\ensuremath{\PW\PW}\xspace}
\newcommand{\tZq}{\ensuremath{\PQt\PZ\PQq}\xspace}
\newcommand{\ttW}{\ensuremath{\ttbar\PW}\xspace}
\newcommand{\ttH}{\ensuremath{\ttbar\PH}\xspace}
\newcommand{\ttG}{\ensuremath{\ttbar\PGg}\xspace}
\newcommand{\Zg}{\ensuremath{\PZ\PGg}\xspace}
\newcommand{\WZG}{\ensuremath{\PW\PZ\PGg}\xspace}
\newcommand{\Wg}{\ensuremath{\PW\PGg}\xspace}
\newcommand{\Xg}{\ensuremath{\PX\PGg}\xspace}
\newcommand{\VVV}{\ensuremath{\PV\PV\PV}\xspace}
\newcommand{\ZZZ}{\ensuremath{\PZ\PZ\PZ}\xspace}
\newcommand{\WZZ}{\ensuremath{\PW\PZ\PZ}\xspace}
\newcommand{\WWZ}{\ensuremath{\PW\PW\PZ}\xspace}
\newcommand{\WWW}{\ensuremath{\PW\PW\PW}\xspace}
\newcommand{\VH}{\ensuremath{\PV\PH}\xspace}

\newcommand{\lepW}{{\HepParticle{\Pell}{\PW}{}}\xspace}
\newcommand{\thirdlep}{{\HepParticle{\Pell}{3}{}}\xspace}
\newcommand{\fourthlep}{{\HepParticle{\Pell}{4}{}}\xspace}
\newcommand{\lepZI}{{\HepParticle{\Pell}{\PZ}{1}}\xspace}
\newcommand{\lepZII}{{\HepParticle{\Pell}{\PZ}{2}}\xspace}
\newcommand{\eee}{\ensuremath{\Pe\Pe\Pe}\xspace}
\newcommand{\eem}{\ensuremath{\Pe\Pe\PGm}\xspace}
\newcommand{\emm}{\ensuremath{\PGm\PGm\Pe}\xspace}
\newcommand{\mme}{\ensuremath{\PGm\PGm\Pe}\xspace}
\newcommand{\mmm}{\ensuremath{\PGm\PGm\PGm}\xspace}

\newcommand{\MATRIX}{\textsc{matrix}\xspace}
\newcommand{\scale}{\ensuremath{\,\text{(scale)}}\xspace}
\newcommand{\PDF}{\ensuremath{\,\text{(PDF)}}\xspace}

\newcommand{\muR}{\ensuremath{\mu_{\mathrm{R}}}\xspace}
\newcommand{\muF}{\ensuremath{\mu_{\mathrm{F}}}\xspace}
\newcommand{\mZ}{\ensuremath{m_{\PZ}}\xspace}

\newcommand{\NLOQCD}{\ensuremath{\mbox{NLO QCD}}\xspace}
\newcommand{\NNLOQCD}{\ensuremath{\mbox{NNLO QCD}}\xspace}
\newcommand{\NNLOQCDNLOEW}{\ensuremath{\mbox{NNLO QCD}\times\mbox{NLO EW}}\xspace}

\newcommand{\Pellpr}{{\HepParticle{\Pell}{}{\prime}}\xspace}
\newcommand{\abseta}{\ensuremath{\abs{\eta}}\xspace}
\newcommand{\absetalepW}{\ensuremath{\abs{\eta(\lepW)}}\xspace}
\newcommand{\Nell}{\ensuremath{N_{\Pell}}\xspace}
\newcommand{\NOSSF}{\ensuremath{N_{\mathrm{OSSF}}}\xspace}
\newcommand{\absmllmZ}{\ensuremath{\abs{m(\lepZI,\lepZII)-\mZ}}\xspace}
\newcommand{\mllmZ}{\ensuremath{m(\lepZI,\lepZII)}\xspace}
\newcommand{\minmll}{\ensuremath{\min(m(\Pell,\Pellpr))}\xspace}
\newcommand{\mll}{\ensuremath{m(\Pell,\Pellpr)}\xspace}
\newcommand{\mlll}{\ensuremath{m(\lepZI,\lepZII,\lepW)}\xspace}

\providecommand{\cmsTable}[1]{\resizebox{\textwidth}{!}{#1}}
\newlength\cmsTabSkip\setlength{\cmsTabSkip}{1ex}

\cmsNoteHeader{SMP-24-005}
\title{Measurement of the inclusive \texorpdfstring{\WZ}{WZ} production cross section in \texorpdfstring{\pp}{pp} collisions at \texorpdfstring{$\sqrt{s}=13.6\TeV$}{sqrt(s)=13.6 TeV}}

\date{\today}

\abstract{The inclusive \WZ production cross section is measured in proton-proton collisions at a centre-of-mass energy of 13.6\TeV, using data collected during 2022 with the CMS detector, corresponding to an integrated luminosity of 34.7\fbinv. The measurement uses multileptonic final states and a simultaneous likelihood fit to the number of events in four different lepton flavour categories: \eee, \eem, \mme, and \mmm. The selection is optimized to minimize the number of background events, and relies on an efficient prompt lepton discrimination strategy. The \WZ production cross section is measured in a phase space defined within a 30\GeV window around the \PZ boson mass, as $\sigma_{\text{total}}(\pp\to\WZ)=55.2\pm1.2\stat\pm1.2\syst\pm0.8\lum\pm0.3\thy\unit{pb}$. In addition, the cross section is measured in a fiducial phase space closer to the detector-level requirements. All the measurements presented in this paper are in agreement with standard model predictions.}

\hypersetup{%
pdfauthor={CMS Collaboration},%
pdftitle={Measurement of the inclusive WZ production cross section in pp collisions at sqrt(s)=13.6 TeV},%
pdfsubject={CMS},%
pdfkeywords={CMS, electroweak, Run 3}}

\maketitle

\section{Introduction}

Diboson production cross sections are interesting physical observables at the high energy proton-proton (\pp) collisions at the CERN LHC. With relatively large event yields, high-purity multileptonic final states, and sensitivity to deviations of the trilinear gauge couplings (TGCs)~\cite{Degrande:2012wf} from their standard-model (SM) predictions, diboson production is one of the best experimental channels to study the properties of the electroweak (EW) sector of the SM.

\begin{figure}[!ht]
\centering
\includegraphics[width=0.31\textwidth]{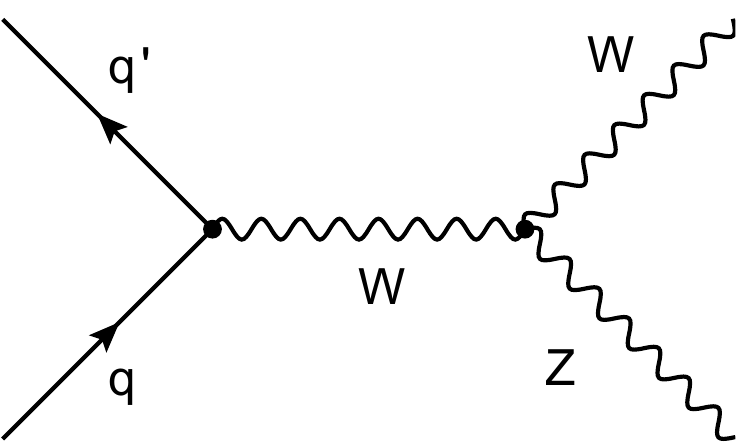}%
\hfill%
\includegraphics[width=0.31\textwidth]{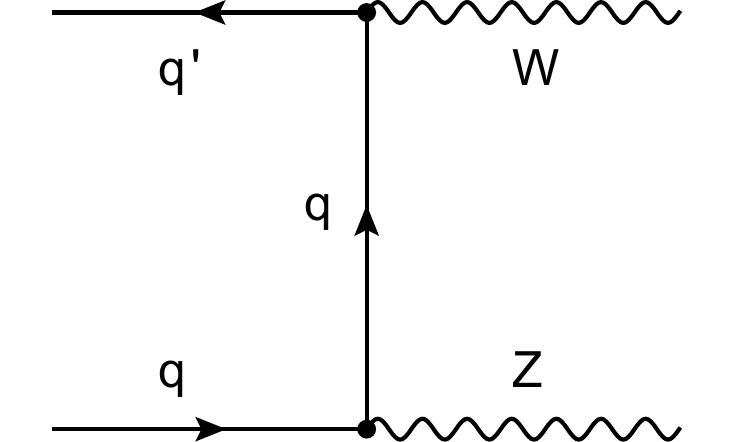}%
\hfill%
\includegraphics[width=0.31\textwidth]{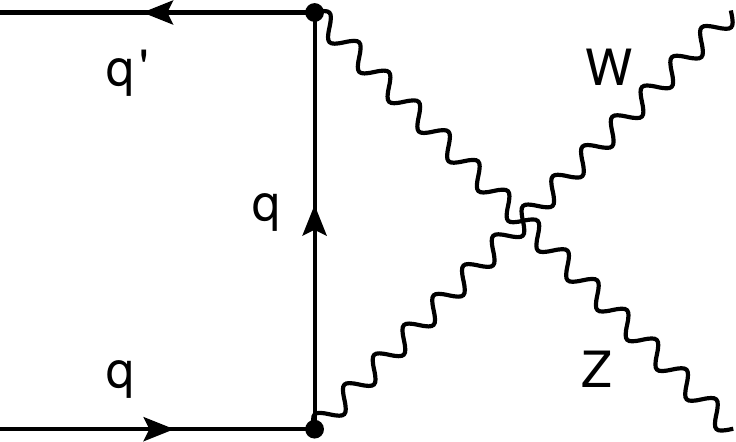}%
\caption{Leading-order Feynman diagrams for \WZ production  in \pp collisions from the
$s$ channel (left), $t$ channel (centre), and $u$ channel (right).}
\label{fig:feynWZ}
\end{figure}

The simultaneous production of a \PW and a \PZ boson (\WZ) in \pp collisions is dominated by quark-antiquark annihilation at tree level, as can be seen in Fig.~\ref{fig:feynWZ}. As a result, the process is sensitive to TGCs that could modify the \WZ production cross section and thus provide evidence of beyond-the-SM (BSM) physics. Additionally, this process is one of the main backgrounds for many SM measurements and BSM searches in multileptonic final states.

The \WZ production cross section was measured at the LHC in \pp collisions at centre-of-mass energies of 5, 7, 8, and 13\TeV~\cite{CMS:2016qth,ATLAS:2016bkj,Khachatryan:2016tgp,Aaboud:2016yus,Aaboud:2019gxl,CMS:2021pqj,CMS:2021icx} by the CMS and ATLAS Collaborations. All these measurements are in good agreement with the most accurate SM predictions within uncertainties. In this paper, a new measurement of the inclusive \WZ production cross section in final states with three leptons that can be either electrons (\Pe) or muons (\PGm) is presented. The analysed data were collected by the CMS experiment in \pp collisions at a centre-of-mass energy of 13.6\TeV during 2022, the first data-taking year of the \mbox{Run 3} of the LHC, and correspond to a total integrated luminosity of 34.7\fbinv. The measurement is performed by classifying events according to the number of leptons of a given flavour. The cross section is determined via a simultaneous maximum likelihood fit to the event counts in all categories, as well as via fits to each category individually.

This paper is organized as follows. Section~\ref{sec:detector} describes the CMS detector. Section~\ref{sec:simul} enumerates the simulated samples used in the measurement. Section~\ref{sec:reconstruction} details the event reconstruction and the data sample used in this analysis. Section~\ref{sec:selection} presents the event selection and signal region definition. Section~\ref{sec:background} explains the main SM background processes and the techniques used for their estimation. Section~\ref{sec:systematics} includes a summary of the sources of uncertainty and their impact on the measurement. Section~\ref{sec:results} presents the measurement strategy for signal extraction and the results. Finally, a summary is presented in Section~\ref{sec:summary}.

Tabulated results are provided in the HEPData record for this analysis~\cite{hepdata}.

\section{The CMS detector}
\label{sec:detector}

The central feature of the CMS apparatus is a superconducting solenoid of 6\unit{m} internal diameter,
providing a magnetic field of 3.8\unit{T}. Within the solenoid volume are a silicon pixel and strip tracker,
a lead tungstate crystal electromagnetic calorimeter (ECAL), and a brass and scintillator hadron calorimeter (HCAL),
each composed of a barrel and two endcap sections. Forward calorimeters extend the pseudorapidity ($\eta$) coverage
provided by the barrel and endcap detectors. Muons are detected in gas-ionization chambers embedded in the steel
flux-return yoke outside the solenoid. A more detailed description of the CMS detector, together with a definition
of the coordinate system used and the relevant kinematic variables, can be found in Refs.~\cite{Chatrchyan:2008zzk, CMS:PRF-21-001}.

Events of interest are selected using a two-tiered trigger system. The first level, composed of custom hardware processors, uses information from the calorimeters and muon detectors to select events at a rate of around 100\unit{kHz} within a fixed latency of about 4\mus~\cite{CMS:2020cmk}. The second level, known as the high-level trigger, consists of a farm of processors running a version of the full event reconstruction software optimized for fast processing, and reduces the event rate to around 5\unit{kHz} before data storage~\cite{CMS:2016ngn, CMS:TRG-19-001}.

\section{Simulated samples}
\label{sec:simul}

Signal \WZ events are generated using \POWHEG \textsc{box} (v2.0)~\cite{Nason:2004rx,Frixione:2007vw,Alioli:2010xd,Melia:2011tj,Nason:2013ydw} at next-to-leading order (NLO) in quantum chromodynamics (QCD), considering only the charged lepton decays of the \WZ system. Renormalization (\muR) and factorization (\muF) scales in the signal prediction are computed dynamically based on the invariant mass of the \WZ system in the generated event, according to the \POWHEG method~\cite{Frixione:2007vw}. The SM background processes are simulated at NLO in QCD using \POWHEG for several diboson processes (\WW, \ZZ) and top quark production processes (\ttbar and single top), at LO in QCD using \MCFM (v7.0)~\cite{Campbell:2010ff} for the gluon-gluon loop induced \ZZ, and with \MGvATNLO~\cite{MADGRAPH5} (v2.9.13) for the remaining diboson processes (\Zg, \Wg, and \VH; where \PV stands for \PW and \PZ), triboson production (\WWW, \WWZ, \WZZ, \ZZZ, \WZG), and associated production of top quarks with other SM particles (\ttW, \ttZ, \ttH, \ttG, \tZq). All the \MGvATNLO processes have been generated at NLO in QCD, considering additional partons in the matrix element wherever possible. In all cases the NNPDF3.1\_nnlo\_hessian\_pdfas~\cite{Ball:2017nwa} parton distribution function (PDF) set is used for the matrix element generation, and the generator is interfaced with \PYTHIA8 (v3.0.6)~\cite{Bierlich:2022pfr} for the showering, hadronization, and underlying-event modelling with the CP5 tune~\cite{Sirunyan_2020}.

The propagation of the generated particles through the CMS detector and the modelling of the detector response are performed using \GEANTfour~\cite{Agostinelli:2002hh} using the alignment and calibration from the collision data. The simulated events are processed with the same reconstruction algorithms used for the collision data.

\section{Event reconstruction}
\label{sec:reconstruction}

The particle-flow (PF) algorithm~\cite{CMS:2017yfk} reconstructs and identifies each individual particle in an event with an optimized combination of information from the various elements of the CMS detector. The energy of photons is obtained from the ECAL measurement. The energy of electrons is determined from a combination of the electron momentum at the primary interaction vertex as determined by the tracker, the energy of the corresponding ECAL cluster, and the energy sum of all bremsstrahlung photons spatially compatible with originating from the electron track. The primary vertex (PV) is taken to be the vertex corresponding to the hardest scattering in the event, evaluated using tracking information alone, as described in Section 9.4.1 of Ref.~\cite{CMS-TDR-15-02}. The energy of muons is obtained from the curvature of the corresponding track, both inside the solenoid and in the return yoke. The energy of charged hadrons is determined from a combination of their momentum measured in the tracker and the matching ECAL and HCAL energy deposits, corrected for the response function of the calorimeters to hadronic showers. Finally, the energy of neutral hadrons is obtained from the corresponding corrected ECAL and HCAL energies.

All electrons in the analysis must have a transverse momentum (\pt) greater than 10\GeV and $\abseta<2.5$. The electron identification is based on a multivariate analysis (MVA) discriminant that combines observables describing the matching of the measurements in the tracker and the ECAL, the description of energy clusters in the ECAL, and the amount of bremsstrahlung radiation emitted during the propagation through the detector. The presence of electrons resulting from asymmetric photon conversions is reduced by requiring that their associated track have no missing hits in the innermost layers of the silicon tracker. An additional criterion that requires consistency among three measurements of the electron charge~\cite{CMS:2020uim,CMS:2015xaf} is applied to ensure a reliable charge determination. Muons are reconstructed by combining the information from the tracking system and the muon spectrometers in a global fit~\cite{CMS:2018rym} that identifies the quality of the matching between the tracker and muon systems, and imposes minimal requirements in the track-related quantities. Muons are required to pass the kinematic selection requirements of $\pt>5\GeV$ and $\abseta<2.4$. As in the case of electrons, an additional identification requirement is imposed to reduce the probability of charge misassignment in muons by requiring the measured curvature to have an uncertainty smaller than 20\%.

An additional set of identification and isolation criteria are applied to define the restricted set of light leptons (\Pe, \PGm) used in the analysis. These additional requirements target a high selection efficiency for leptons from \PW or \PZ bosons or from \PGt lepton decays, while rejecting those coming from other sources, such as leptons from hadron decays (non-prompt leptons). Lepton isolation requirements are imposed following the same approach as in Ref.~\cite{CMS:2021icx}. The isolation of each lepton is defined as the scalar \pt sum of all photons and charged and neutral hadrons in a cone around the lepton direction. The contributions from neutral particles originating from additional \pp interactions, within the same or nearby bunch crossings (pileup) are estimated based on their average spatial energy density and the effective area covered by an isolation cone that is placed around the position of a given neutral particle in the detector. Charged particles identified to be originating from pileup vertices are discarded for this computation. A requirement on the lepton isolation relative to its \pt to be smaller than 0.4 is applied. Leptons that pass the above identification and isolation criteria are referred to as ``loose'' leptons.

An MVA discriminant that aims to separate the prompt and non-prompt contributions is further used to select high-quality prompt leptons~\cite{CMS:2023dsu}. Its inputs are the charged and neutral components of the lepton isolation as previously described, the properties of particles reconstructed in close proximity to the lepton, and the impact parameter with respect to the PV of the reconstructed lepton track in two and three dimensions. These selection criteria are more relevant for electrons and hence the requirements on the MVA discriminant are tighter than for muons.

For each event, hadronic jets are clustered from the PF candidates using the anti-\kt algorithm~\cite{Cacciari:2008gp, Cacciari:2011ma} with a distance parameter of 0.4. Jet momentum is determined as the vectorial sum of all particle momenta in the jet, and is found from simulation to be, on average, within 5 to 10\% of the true momentum over the whole \pt spectrum and detector acceptance. Pileup interactions can contribute additional tracks and calorimetric energy deposits, increasing the apparent jet momentum. The pileup-per-particle identification algorithm~\cite{Sirunyan:2020foa,Bertolini:2014bba} is used to mitigate the effect of pileup, making use of local shape information, event pileup properties, and tracking information. For each neutral particle, a local shape variable is computed using the surrounding charged particles compatible with the PV within $\abseta<2.5$, and using both charged and neutral particles in the region outside of the tracker coverage. This variable is used to distinguish between collinear and soft diffuse distributions of other objects surrounding the particle under consideration. Charged particles identified to be originating from pileup vertices are discarded. The momenta of the neutral particles are then rescaled according to their probability to originate from the PV deduced from the local shape variable, superseding the need for jet-based pileup corrections. Jets produced by the hadronization of \PQb quarks with $\pt>25\GeV$ and $\abseta<2.5$ are identified using the \textsc{DeepJet} algorithm~\cite{BTV-16-002,Bols:2020bkb}, which is based on a deep neural network that combines information from the reconstructed tracks and secondary vertices associated with the jet. The working point used in this analysis corresponds to a \PQb tagging efficiency of 70\% and a mistagging rate of 1\% for light-flavoured jets and gluon jets.

The missing transverse momentum is computed as the negative sum of the \pt of all the PF candidates weighted by their probability to originate from the PV~\cite{CMS:2019ctu}, and its magnitude is denoted as \ptmiss. The \ptmiss is modified to account for corrections to the energy scale of the reconstructed jets in the event. Anomalous high-\ptmiss events can be due to a variety of reconstruction failures, detector malfunctions, or noncollision backgrounds. Such events are rejected by event filters designed to identify more than 85--90\% of the spurious high-\ptmiss events with a mistagging rate smaller than 0.1\%~\cite{CMS-DP-2020-018}.

\section{Event selection}
\label{sec:selection}

The data are recorded with a combination of single- and double-lepton triggers requiring the presence of one or two isolated electrons or muons with loose identification and isolation criteria. The \pt requirement of the single-lepton trigger is 32\GeV for electrons and 24\GeV for muons. The double-lepton triggers require the leading lepton to have a \pt greater than 23 (18)\GeV, and the subleading lepton to have a \pt greater than 12 (8)\GeV for electrons (muons).

Events are required to fulfil strict requirements designed to achieve a pure \WZ signal region (SR) while keeping a high signal efficiency. The selection of events entering the SR is based on the presence of exactly three isolated leptons, with at least two leptons with electric charges of opposite sign and same flavour (OSSF). In order to exploit the kinematic properties of the \WZ production, a simple assignment algorithm is used to tag each of the leptons originating from the \PW and \PZ boson decays. The algorithm works as follows: if only one OSSF lepton pair is found, the leptons are assumed to arise from the \PZ boson decay and, therefore, are labelled as \lepZI and \lepZII. In case two OSSF lepton pairs are found, the leptons of the pair with dilepton invariant mass closest to the nominal \PZ boson mass are selected as \lepZI and \lepZII. In either case, the hardest lepton not associated with the best \PZ boson candidate is referred to as \lepW. After assigning the origin of the leptons, the following additional \pt requirements are imposed in a consistent manner with the trigger requirements: $\pt(\lepZI)>25\GeV$, $\pt(\lepZII)>15\GeV$, and $\pt(\lepW)>25\GeV$. After this minimal preselection, \lepZI and \lepZII are properly tagged to the corresponding \PZ boson with 97\% efficiency, and \lepW is found to be properly tagged to the \PW boson (when present) with a 94\% efficiency. On top of this baseline selection, four channels based on the possible flavour combinations of the tagged leptons (\lepZI, \lepZII, \lepW): \eee, \eem, \mme, and \mmm are considered for the classification of events.

To increase the purity of \WZ events, we require that the invariant mass of the \lepZI and \lepZII to be consistent with the \PZ boson mass $\absmllmZ<15\GeV$, which removes contributions from nonresonant processes. The value of \mZ is taken from the \PZ mass world average given by the PDG~\cite{ALEPH:2005ab}. A minimal requirement of $\ptmiss>35\GeV$ is included to reduce the presence of dileptonic \PZ production with an associated jet that can be misidentified as a lepton. Events with at least one \PQb-tagged jets are vetoed from the SR in order to reduce contributions from top quark processes. Infrared safety in the same-flavour final states is ensured by rejecting events containing an OSSF lepton pair with invariant mass $\mll<4\GeV$. The invariant mass of the three lepton system is required to be $\mlll>100\GeV$, effectively rejecting most of the contribution from \Zg production in which the photon undergoes asymmetric conversion producing an additional final-state electron. All these requirements define the SR.

Several control regions (CRs) are defined to estimate the contribution of the main background processes. Each CR is designed to be dominated by one background process and is used to estimate its normalization including the CR into the analysis likelihood fit. The CR definitions are obtained by inverting SR requirements, effectively increasing the purity of a given background without selecting events that pass the SR selection. In particular, the \ZZ CR is defined by inverting the fourth-lepton veto and dropping the $\ptmiss>35\GeV$ requirement from the SR definition. The hardest lepton not belonging to the best OSSF pair is then refered to as \thirdlep instead of \lepW in this CR, yet the requirements that are imposed are the same as the ones considered in the case with three leptons. The fourth lepton is also required to pass tight identification criteria and to have $\pt>15\GeV$. The \ttZ CR is obtained by inverting the \PQb-tagged jet veto. Finally, the conversions CR (\Xg CR) is defined by removing the $\absmllmZ<15\GeV$ requirement and inverting the trilepton invariant mass requirement to $\mlll<100\GeV$. Both modifications increase the presence of the process in which a photon is radiated by one of the final-state leptons in dileptonic \PZ boson decays, producing a typical trilepton resonance around the \PZ boson peak. To further increase the purity of \Zg events in this phase space, \ptmiss is required to be smaller than 35\GeV. A summary of the definition of the SR and CRs can be found in Table~\ref{tab:SRCR}.

\begin{table}[ht!]
\centering
\topcaption{Requirements for the definition of the signal and control regions of the analysis. Objects in parentheses relate to the \ZZ CR.}
\label{tab:SRCR}
\renewcommand{\arraystretch}{1.1}
\cmsTable{\begin{tabular}{ccccccccc}
    Region & \Nell & $\pt\{\lepZI,\lepZII,\lepW(\thirdlep),(\fourthlep)\}$ & \NOSSF & \absmllmZ & \ptmiss & $N_{\PQb\,\text{tag}}$ & \minmll & $m(\lepZI,\lepZII,\lepW(\thirdlep))$ \\
    & & ({\GeVns})  & & ({\GeVns}) & ({\GeVns})  & & ({\GeVns}) & ({\GeVns}) \\\hline
    SR & $=$3 & ${>}\{25,15,25\}$ & $\geq$1 & $<$15 & $>$35 & $=$0 & $>$4 & $>$100 \\
    \ZZ CR & $=$4 & ${>}\{25,15,25,15\}$ & $\geq$1 & $<$15 & \NA & $=$0 & $>$4 & $>$100 \\
    \ttZ CR & $=$3 & ${>}\{25,15,25\}$ & $\geq$1 & $<$15 & $>$35 & $>$0 & $>$4 & $>$100 \\
    \Xg CR & $=$3 & ${>}\{25,15,25\}$ & $\geq$1 & \NA & $\leq$35 & $=$0 & $>$4 & $<$100 \\
\end{tabular}}
\end{table}

\section{Background estimation}
\label{sec:background}

The background processes in the SR can be grouped into two different categories. The first category includes the reducible backgrounds that originate mainly due to the presence of non-prompt leptons, as defined in Section~\ref{sec:reconstruction}, typically arising from {\PZ}+jets or \ttbar production in which either a jet is misidentified as a lepton or a lepton is produced in the decay of a heavy-flavour hadron. The second category groups all the irreducible backgrounds including SM processes that naturally produce three or more leptons, such as boson pair production, triboson production, and associated production of top quarks with a \PW, \PZ, or \PH boson.

\subsection{Reducible backgrounds}

The SM processes with non-prompt leptons can have a contribution in the SR through the misidentification of a lepton that does not follow the prompt definition detailed in Section~\ref{sec:reconstruction}, yet satisfies the tight identification requirements applied on the leptons that are used in the analysis. In final states with three leptons, these contributions are mainly attributed to {\PZ}+jets events in which in which a jet from ISR is misidentified as a lepton, and dileptonic \ttbar or single top events where an additional lepton is produced in the in-flight decay of a hadron. The lepton identification criteria have been optimized to reduce the non-prompt rate to roughly 3\% of the total yield in the SR. Due to the relatively low contribution in the final selection, the non-prompt background is modelled considering the shapes and normalizations directly from MC simulation.

This MC-based non-prompt estimation is validated in a sideband region enriched in non-prompt background. This region is characterized by the presence of three final state leptons with kinematic requirements similar to the ones used in the SR, but considering events in which the tight identification criteria defined in Section~\ref{sec:reconstruction} are dropped for \lepW. The majority of non-prompt leptons contaminating the SR are tagged as \lepW  according to the tagging algorithm described in Section~\ref{sec:selection}. Based on this premise, dropping the tight requirement on this lepton enriches the presence of the type of non-prompt backgrounds that need to be controlled in the analysis. The non-prompt purity, expressed as a percentage over the total number of selected events, is 72\% (98\% of it coming from {\PZ}+jets), and  the remaining 28\% can be broken down into a 15\% of \WZ events and 13\% of other prompt processes. The ratio of data and MC was observed to be 0.93 in this region. We further tested the data to MC agreement for several lepton kinematic observables of \lepZI, \lepZII, and \lepW.  Overall agreement was observed for both \lepZI, \lepZII, and a 30\% normalization uncertainty was derived based on these studies to cover normalization effects that were observed particularly at low \pt (\lepW) and high \absetalepW. Other combinations of the non-prompt candidate was studied by dropping the tight criteria on either \lepZI and \lepZII, instead of \lepW, and a similar conclusion was reached. 

\subsection{Irreducible backgrounds}

The main irreducible backgrounds are \ZZ production, amounting to $\sim$5\% of the total yield in the SR, \ttZ production, that contributes $\sim$2\% of the yield in the SR, and the associated production of a photon and other SM particles, \Xg, that represents 1\% of the selected events. 

\begin{figure}[!p]
\centering
\includegraphics[width=0.48\textwidth]{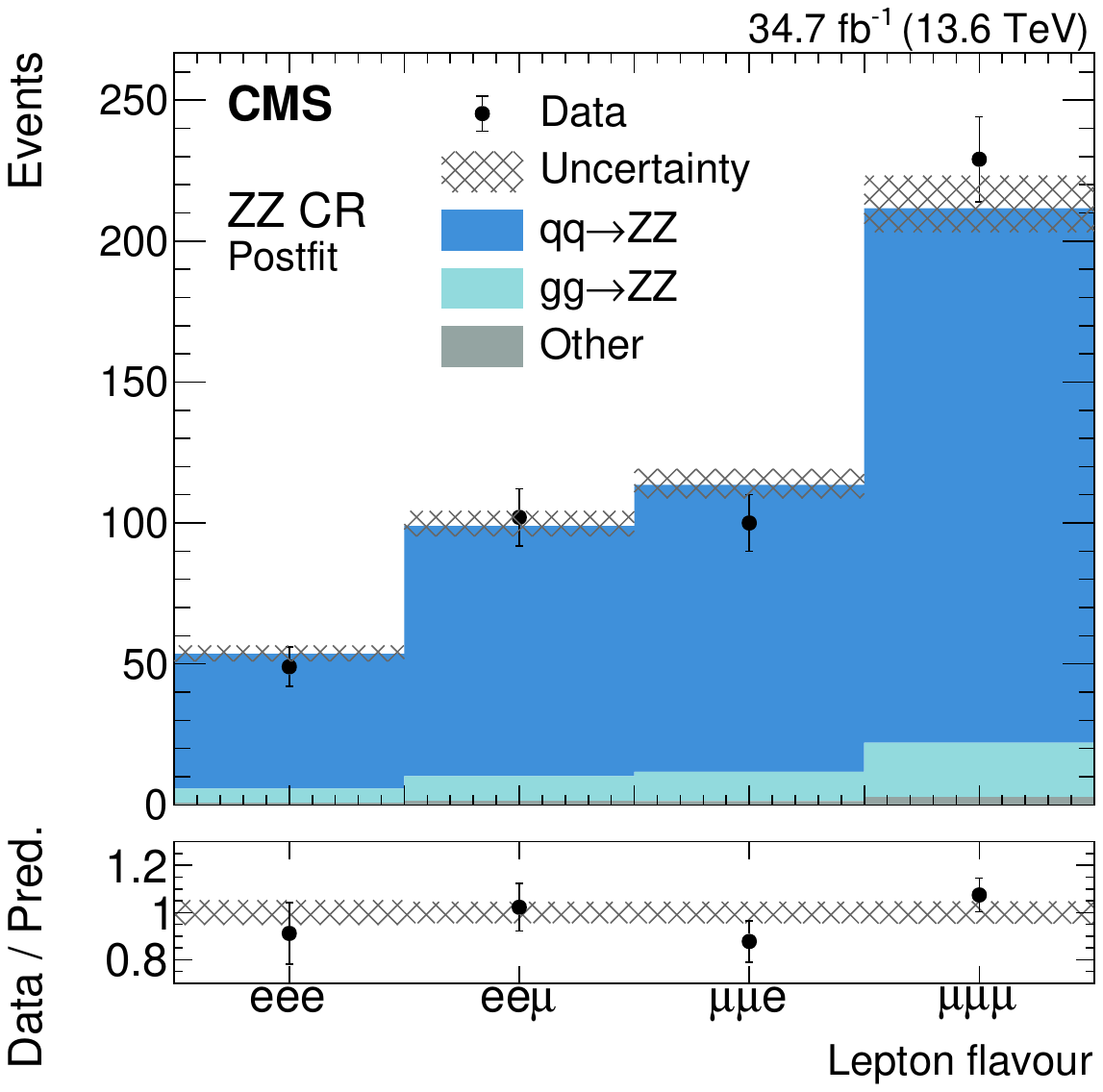}%
\hfill%
\includegraphics[width=0.48\textwidth]{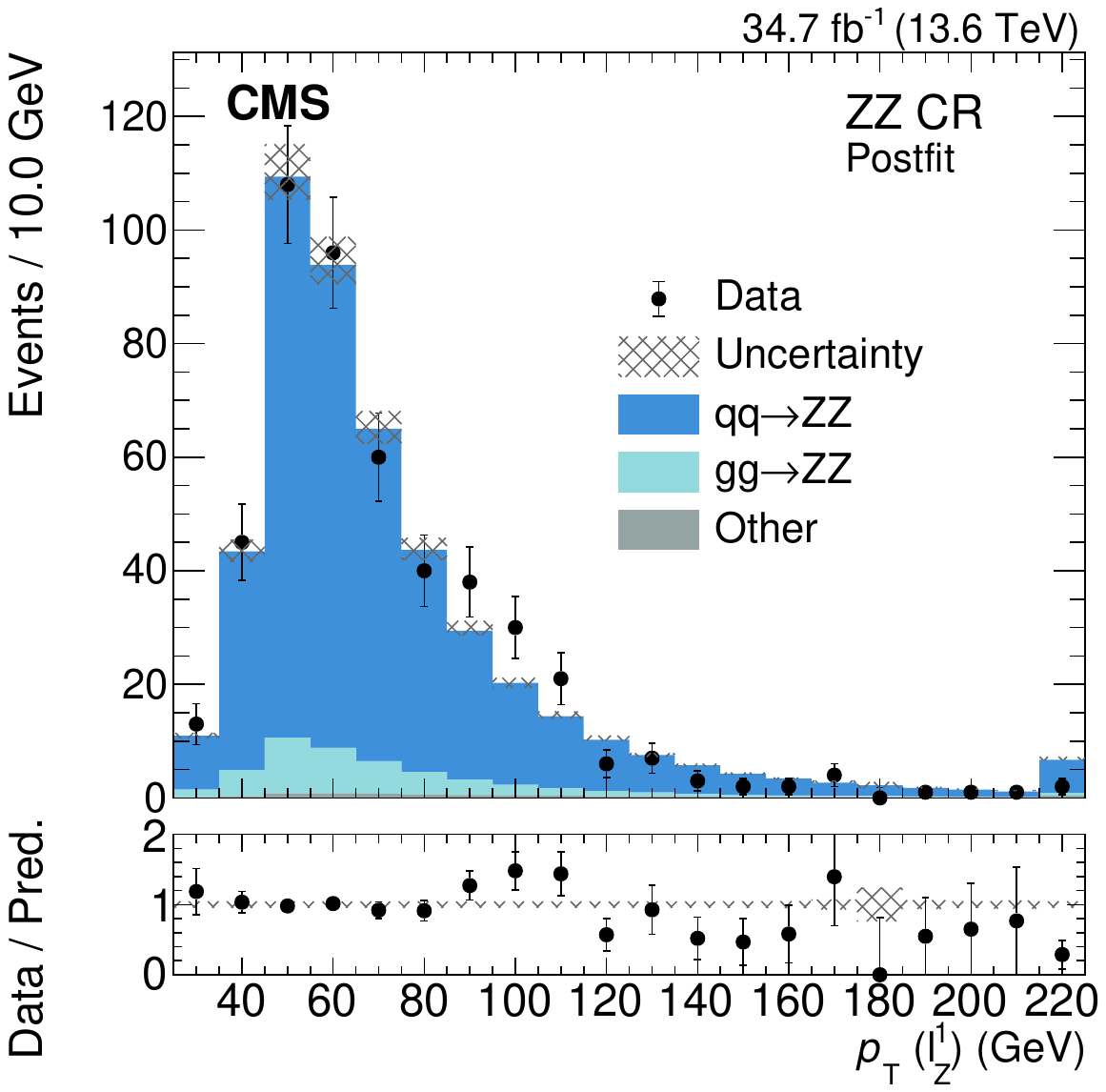} \\[1ex]
\includegraphics[width=0.48\textwidth]{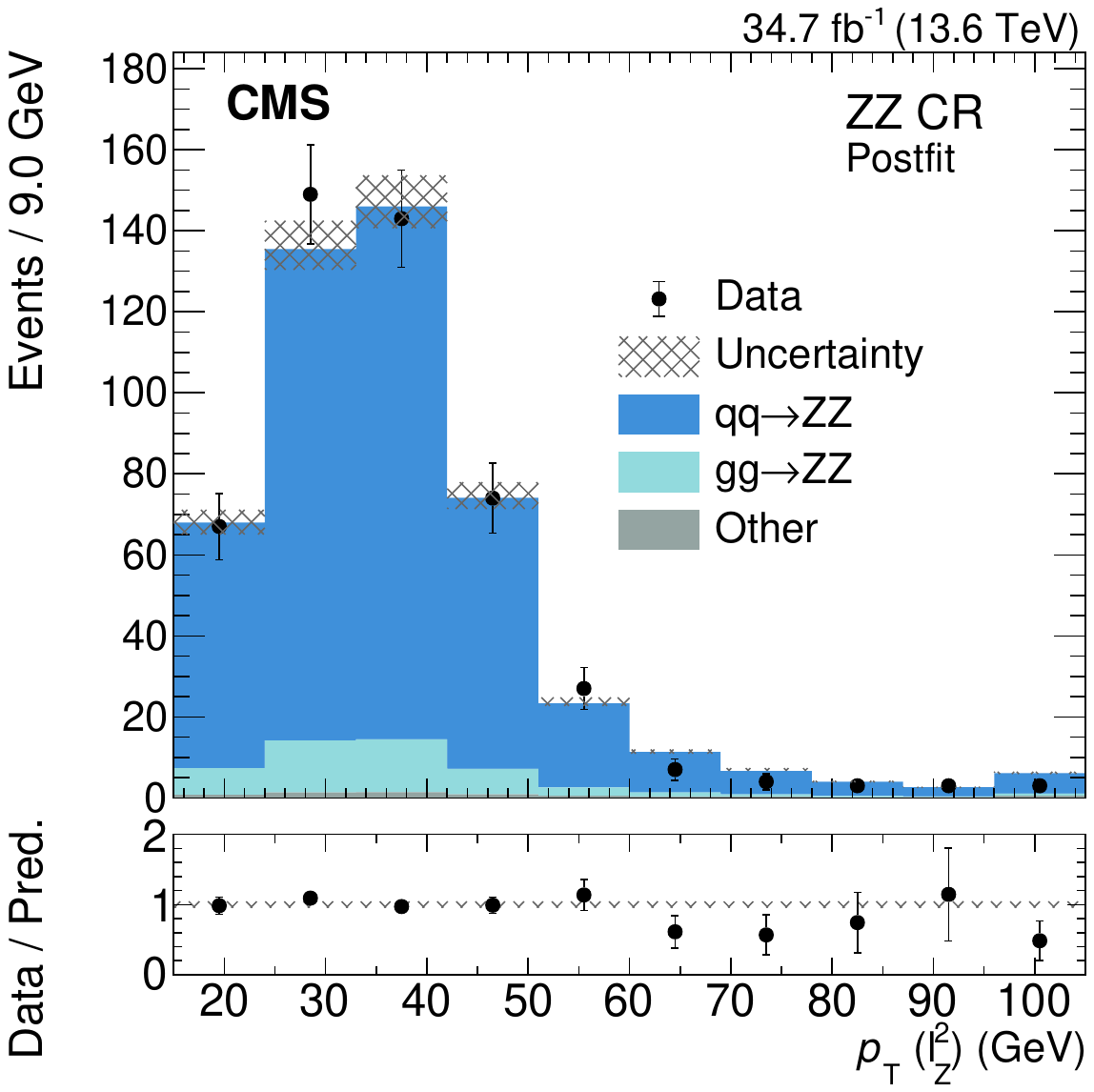}%
\hfill%
\includegraphics[width=0.48\textwidth]{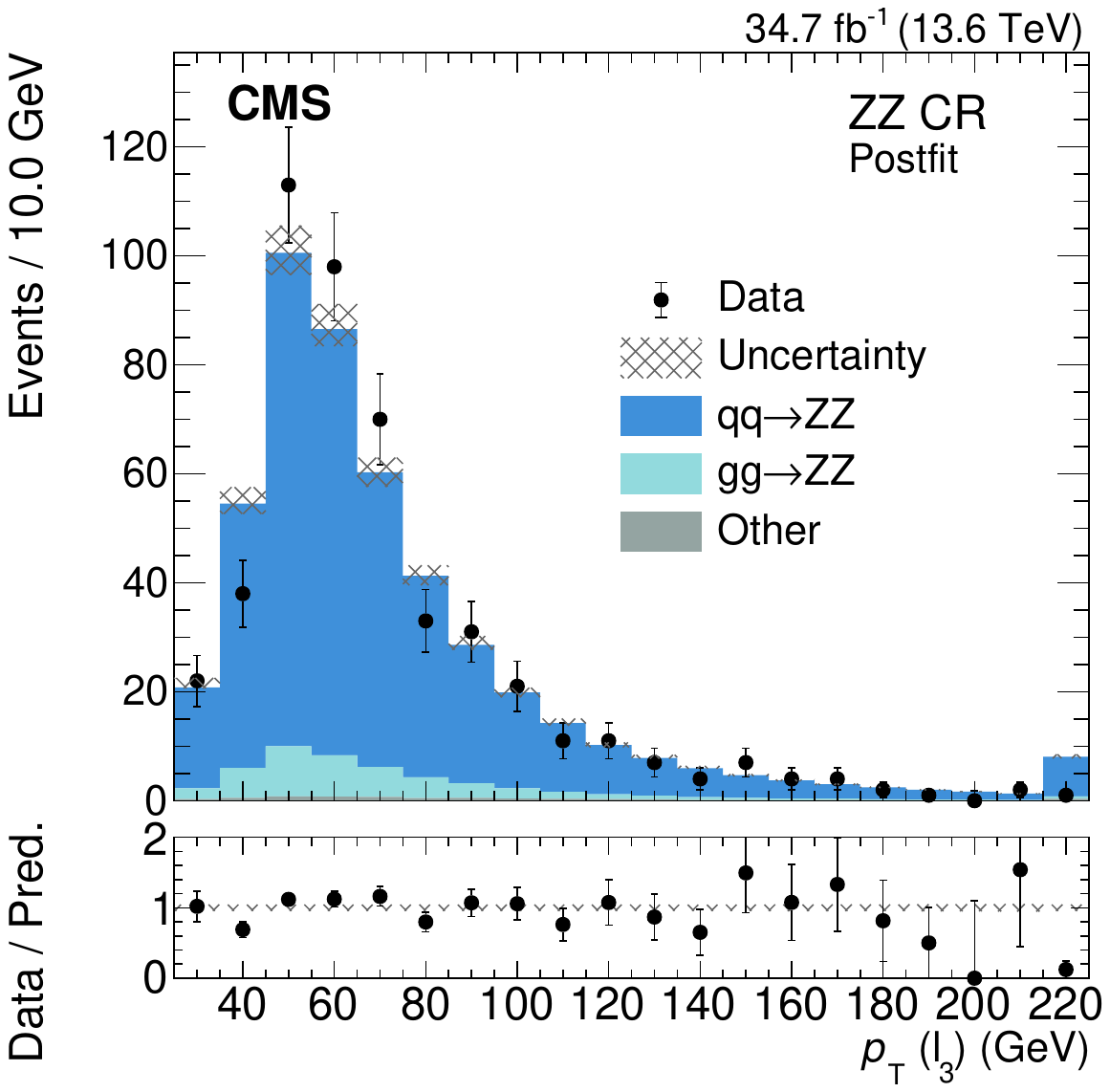}
\caption{Distribution of observables in the \ZZ CR accounting for the fit to data, described in Section~\ref{sec:results}. From upper left (flavour composition), upper right (\pt of \lepZI), lower left (\pt of \lepZII), and lower right (\pt of \thirdlep). The hatched bands show the total uncertainty in the MC prediction. The vertical bars of the data account for the statistical uncertainty. When present, overflow events are included in the last bin of the observables. The ratio panels show the ratio between data (black markers) with respect to the total prediction after the fit to data. Processes with a small contribution to this region are grouped in the ``Other'' category. }
\label{fig:CRZZ}
\end{figure}

\begin{figure}[!tp]
\centering
\includegraphics[width=0.48\textwidth]{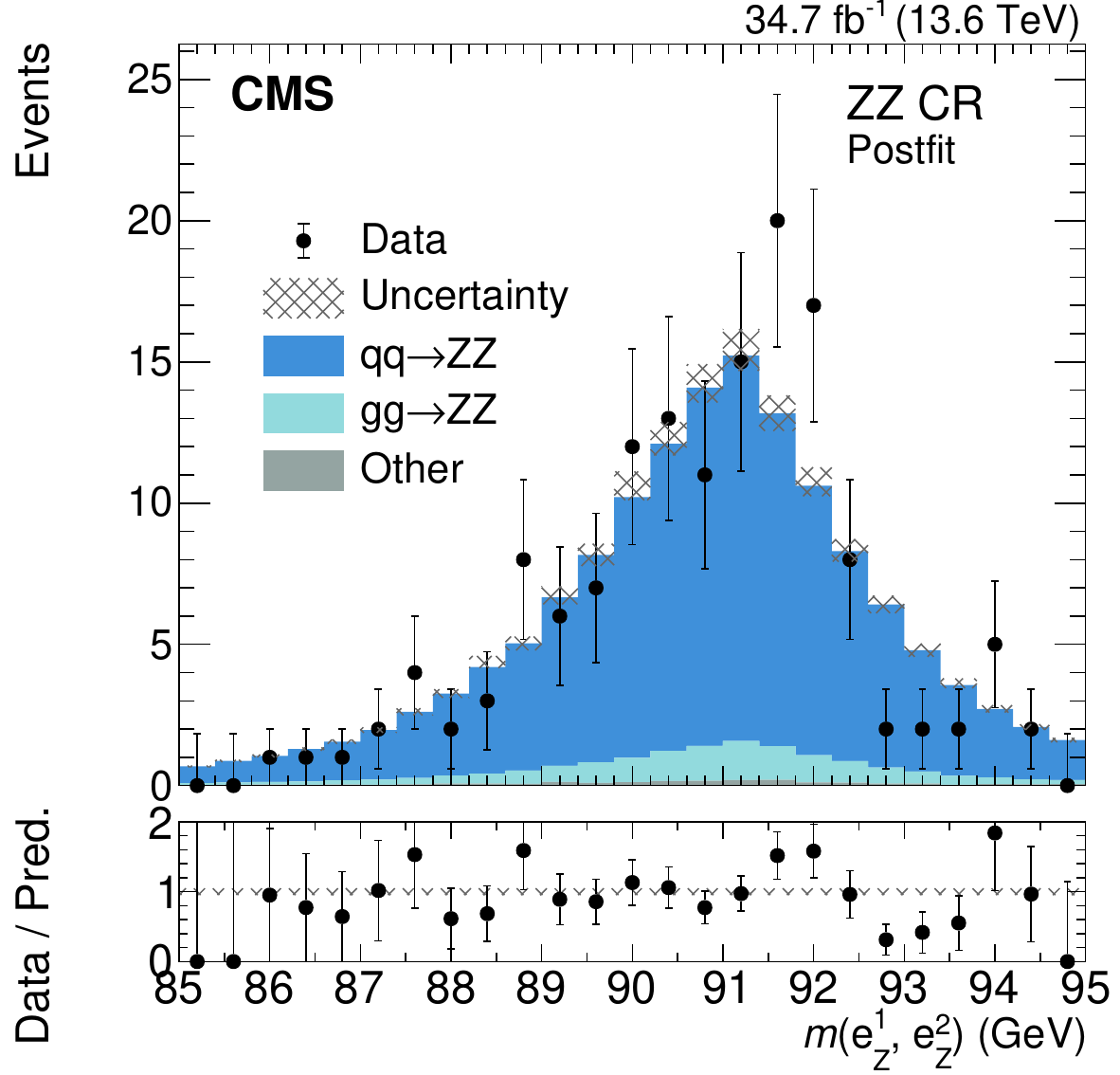}%
\hfill%
\includegraphics[width=0.48\textwidth]{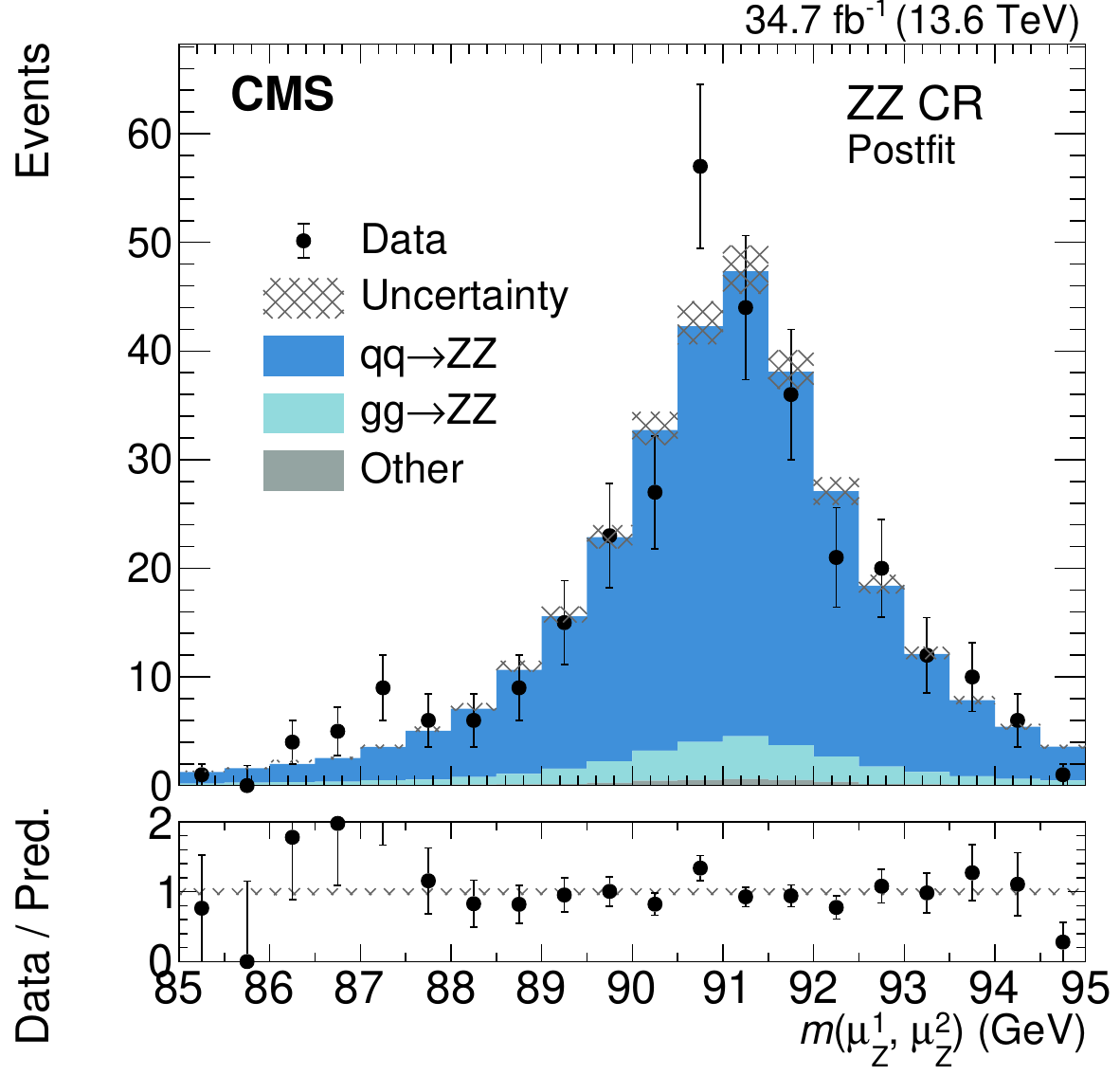}
\caption{Distribution of the invariant mass of \lepZI and \lepZII in the \ZZ CR accounting for the fit to data. The left (right) distribution shows the case in which the leptons are electrons (muons). The hatched bands show the total uncertainty in the MC prediction. The vertical bars of the data account for the statistical uncertainty. The ratio panels show the ratio between data (black markers) with respect to the total prediction after the fit to data. Processes with a small contribution to this region are grouped in the ``Other'' category.}
\label{fig:CRZZ_mll}
\end{figure}

The purity of the \ZZ contribution in the \ZZ CR is about 98\% of the total yield, with 90\% of this contribution coming from \qqbar annihilation (\qqZZ), and the remaining 10\% coming from gluon fusion mechanisms (\ggZZ). The distributions of several observables in this CR are shown in Fig.~\ref{fig:CRZZ}. Distributions of the reconstructed \PZ boson peak considering the two leptons tagged as \lepZI and \lepZII are also shown separately for electrons and muons in Fig.~\ref{fig:CRZZ_mll}. All distributions show relatively good modelling of the \ZZ background.

\begin{figure}[!p]
\centering
\includegraphics[width=0.48\textwidth]{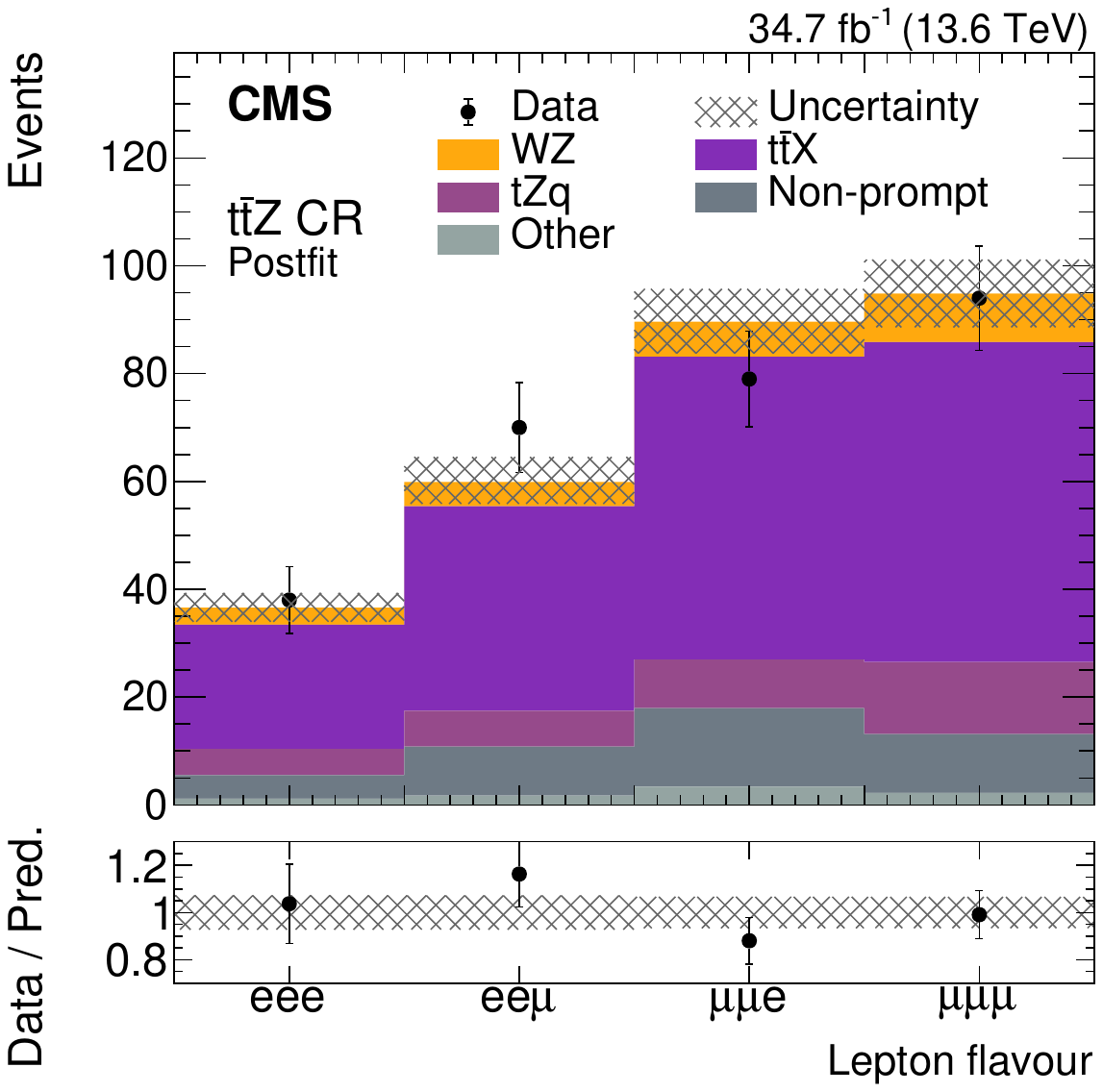}%
\hfill%
\includegraphics[width=0.48\textwidth]{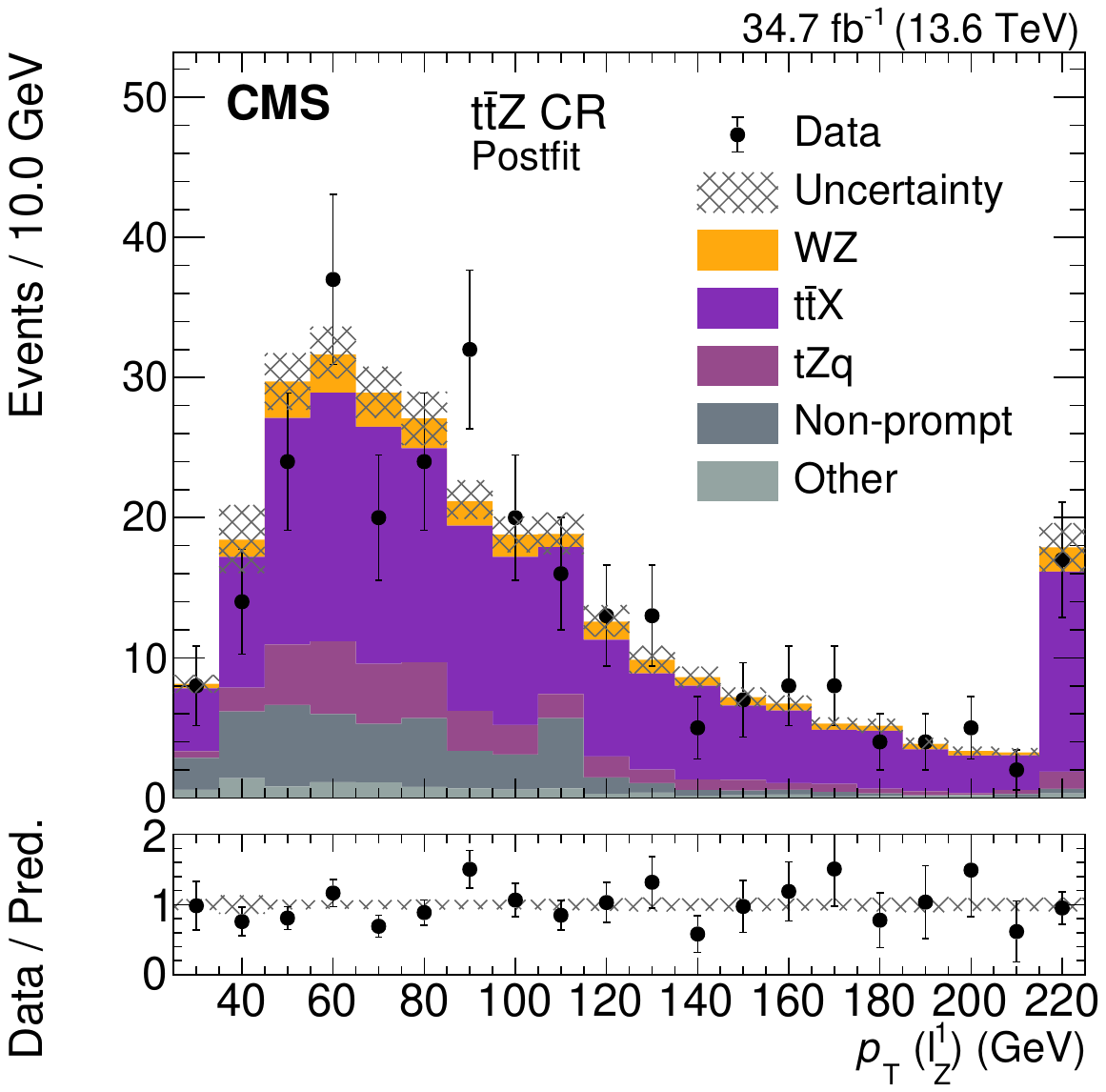} \\[1ex]
\includegraphics[width=0.48\textwidth]{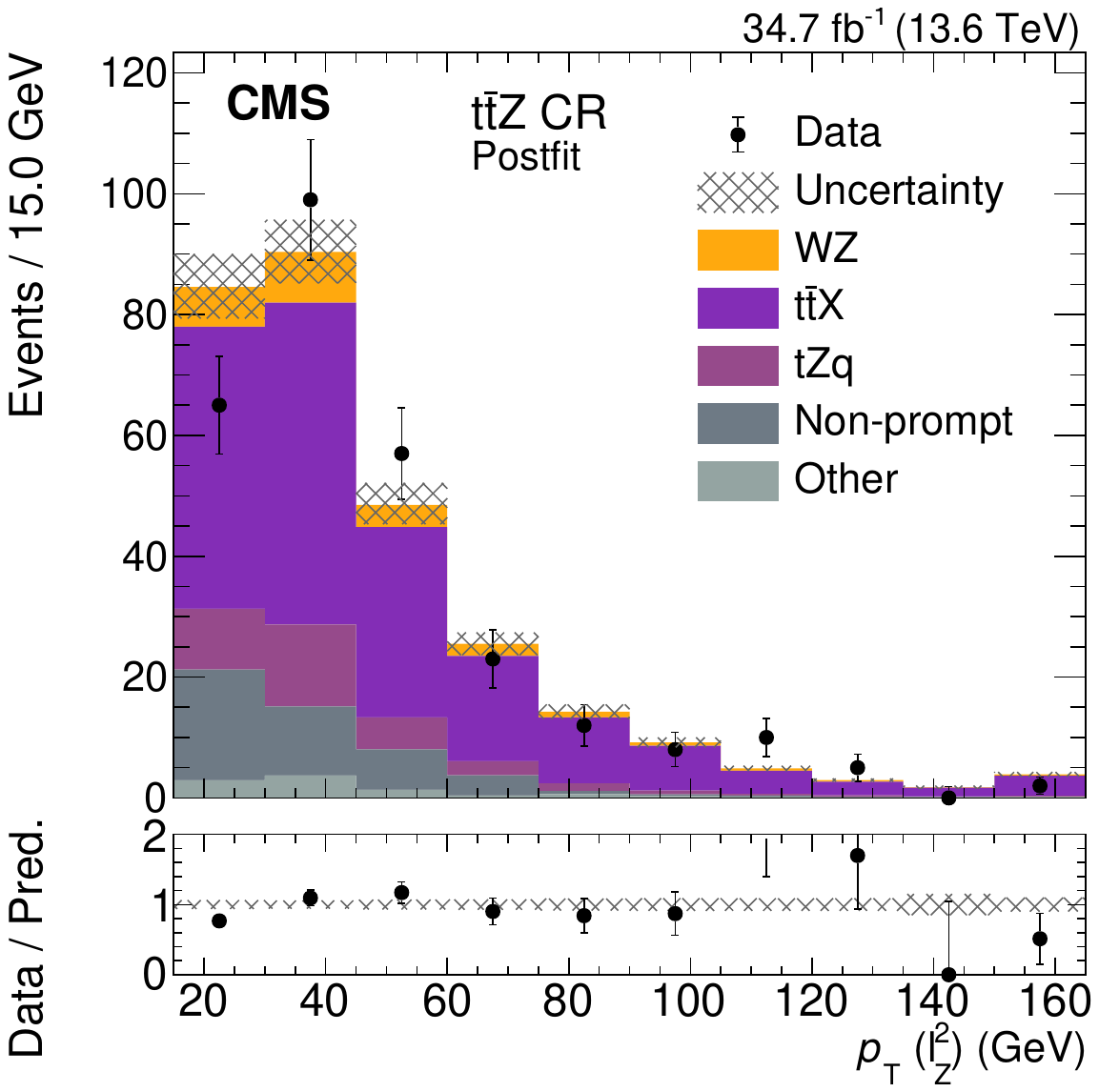}%
\hfill%
\includegraphics[width=0.48\textwidth]{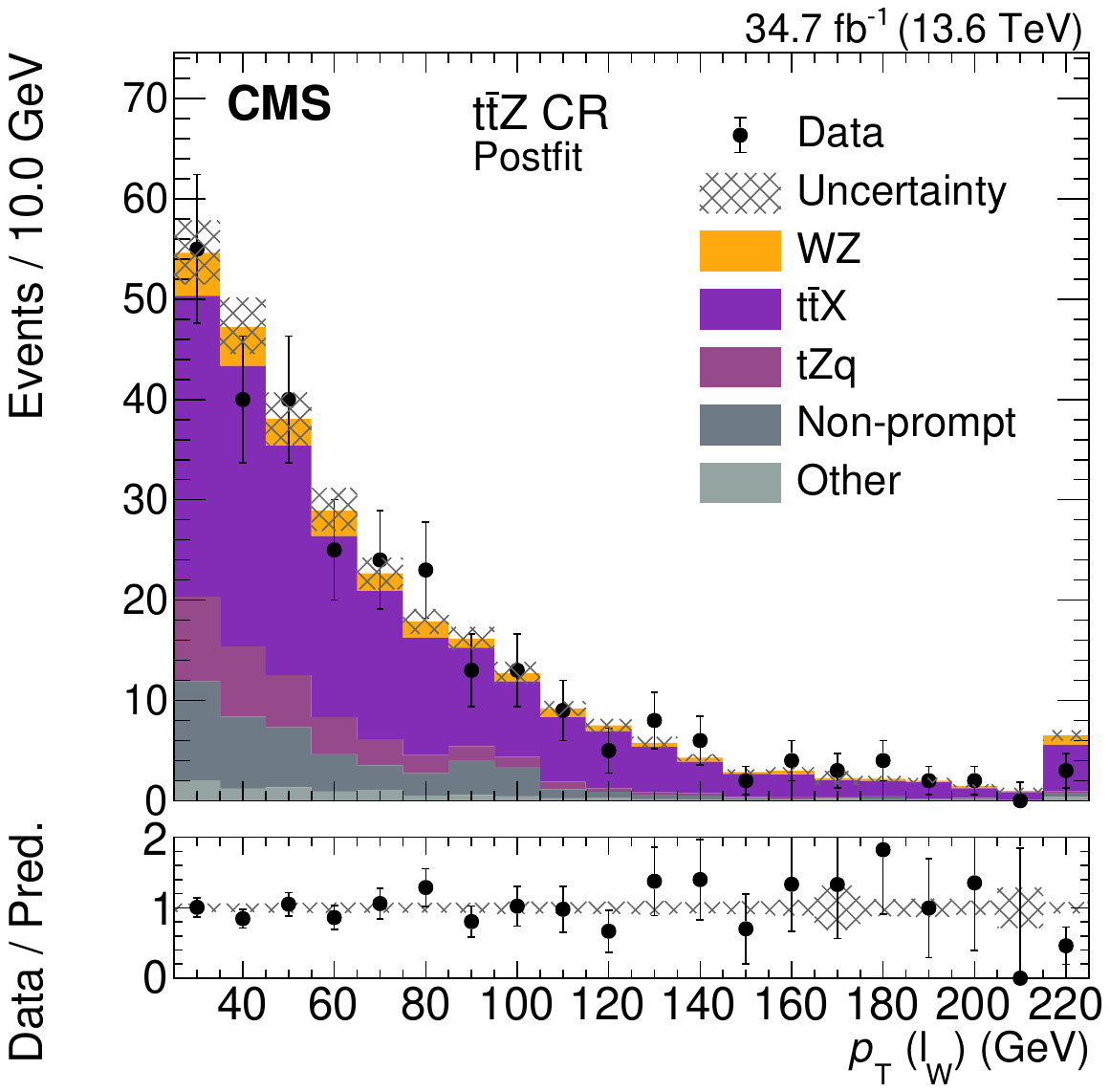}
\caption{Distribution of observables in the \ttZ CR accounting for the fit to data, described in Section~\ref{sec:results}. From upper left (flavour composition), upper right (\pt of \lepZI), lower left (\pt of \lepZII), and lower right (\pt of \lepW). The hatched bands show the total uncertainty in the MC prediction. The vertical bars of the data account for the statistical uncertainty. When present, overflow events are included in the last bin of the observables. The ratio panels show the ratio between data (black markers) with respect to the total prediction after the fit to data. Processes with a small contribution to this region are grouped in the ``Other'' category. }
\label{fig:CRTT}
\end{figure}

The \ttZ CR is especially sensitive to both \ttZ and \tZq production. Residual contributions from \ttW and \ttH production are a factor of 20 smaller than the \ttZ contribution and are grouped with \ttZ into a group labelled \ttX. Good agreement between the observed data and predictions from simulation in this region is demonstrated in Fig.~\ref{fig:CRTT} for several variables.

\begin{figure}[!p]
\centering
\includegraphics[width=0.48\textwidth]{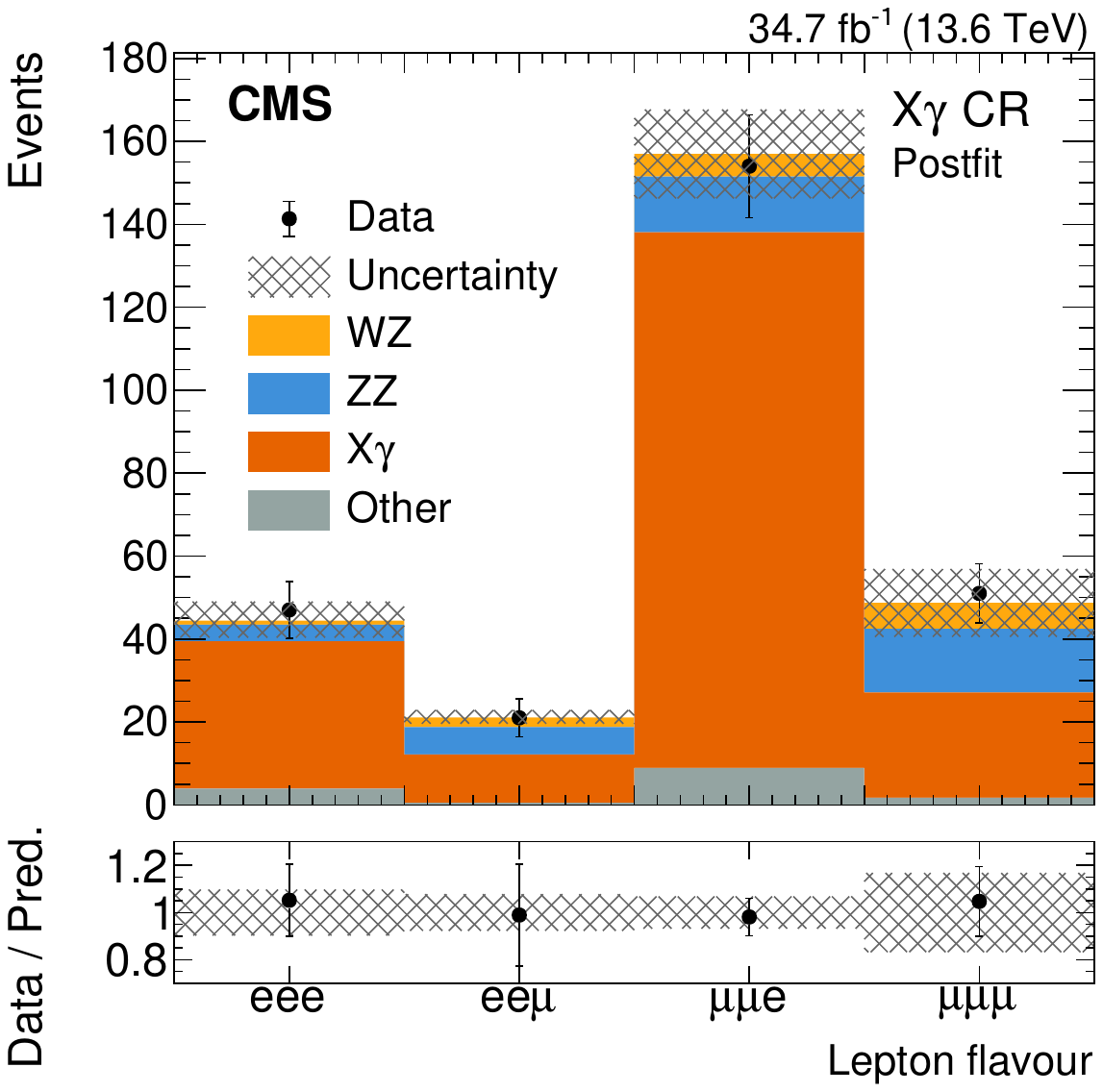}%
\hfill%
\includegraphics[width=0.48\textwidth]{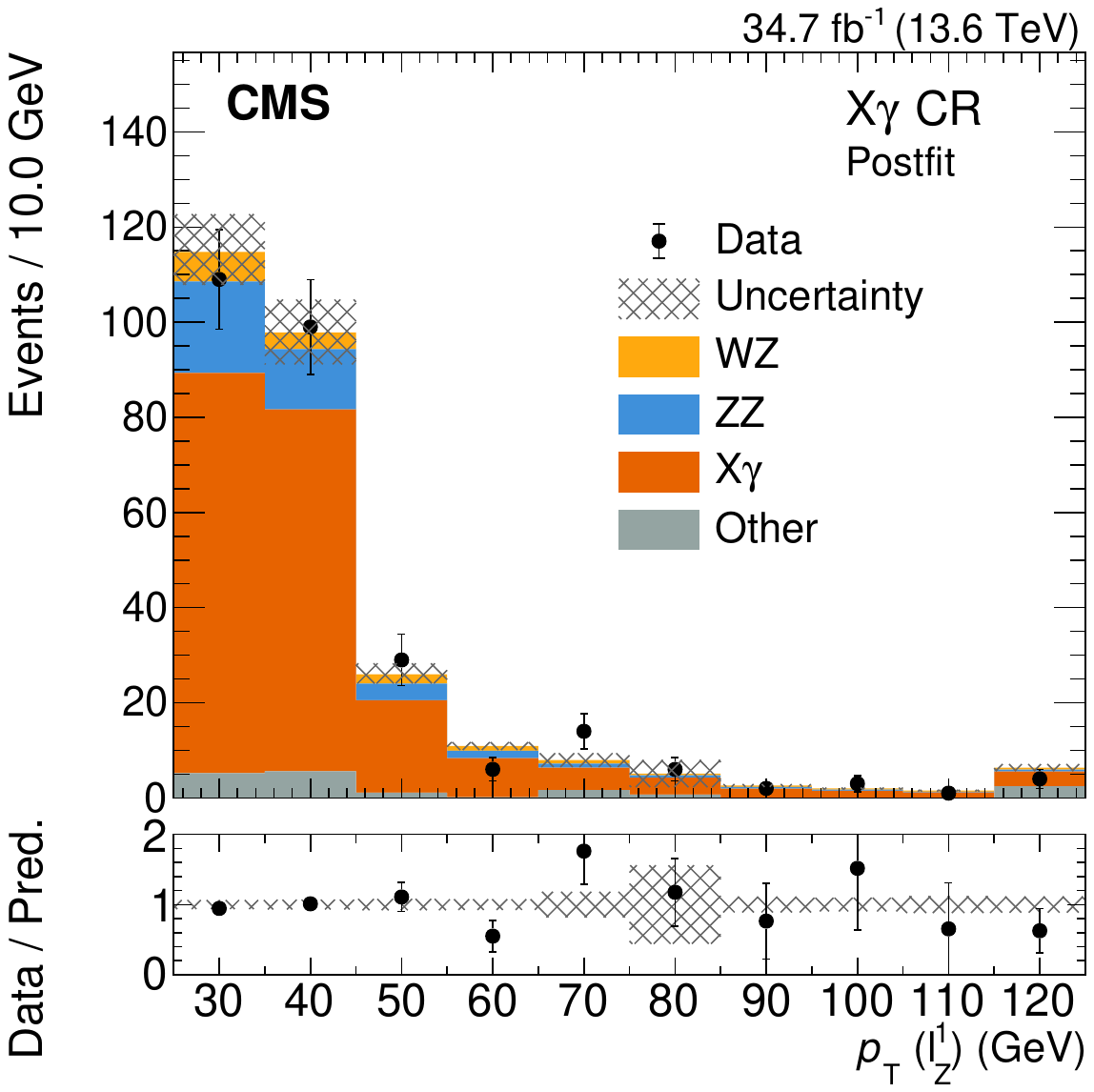} \\[1ex]
\includegraphics[width=0.48\textwidth]{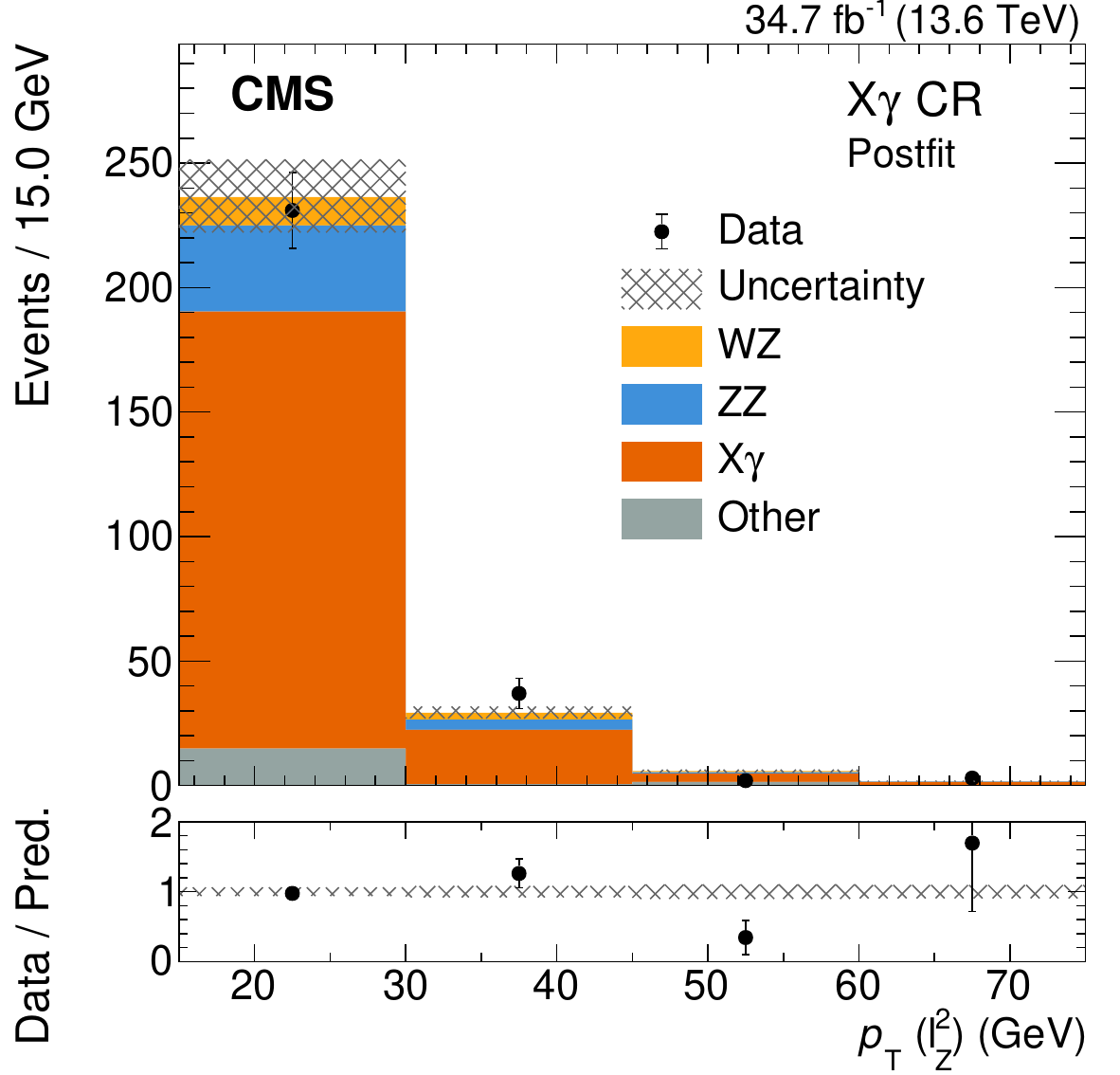}%
\hfill%
\includegraphics[width=0.48\textwidth]{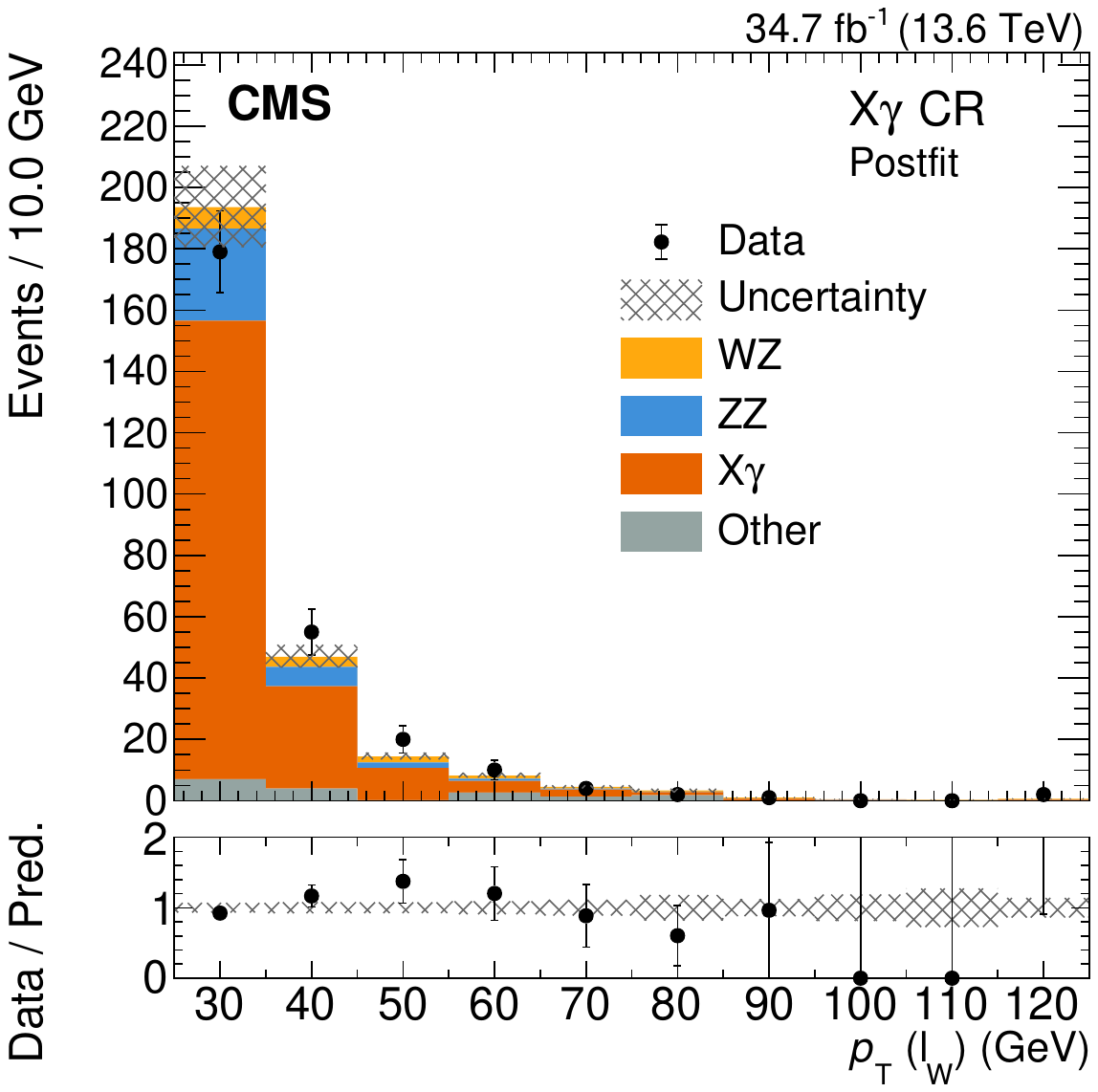}
\caption{Distribution of observables in the \Xg CR accounting for the fit to data, described in Section~\ref{sec:results}. From upper left (flavour composition), upper right (\pt of \lepZI), lower left (\pt of \lepZII), and lower right (\pt of \lepW).  The hatched bands show the total uncertainty in the MC prediction. The vertical bars of the data account for the statistical uncertainty. When present, overflow events are included in the last bin of the observables. The ratio panels show the ratio between data (black markers) with respect to the total prediction after the fit to data. Processes with a small contribution to this region are grouped in the ``Other'' category. }
\label{fig:CRCONV}
\end{figure}

The \Xg background includes processes with a prompt photon that proceeds through an asymmetric conversion due to the interaction with the detector to produce a single detected final-state lepton, or with a virtual photon that converts into a lepton pair. This background is dominated by associated \Zg production (around 99\% of the expected events), with residual contributions from \Wg, \ttG, and \WZG production. The distributions of several relevant observables in the \Xg CR are shown in Fig.~\ref{fig:CRCONV}. Good agreement is also observed in this CR.

The normalization of each major background is considered as an unconstrained parameter in the fit to the data, and the shape is obtained from the simulation. To constrain the major backgrounds, the number of events in flavour categories from each of the CRs is included in the likelihood fit. All the uncertainties included in the total inclusive cross section measurement described in Section~\ref{sec:systematics} also apply to these CRs.

Other irreducible backgrounds contribute less than 1\% to the yields in the SR and are directly estimated from simulation. These include the associated production of a vector boson and a Higgs boson (\VH) and the triboson production (\VVV). These backgrounds are estimated using MC simulation. Contributions from \WZ produced through vector boson scattering have not been considered. This background is expected to contribute as 1\% of the total yield in the SR.

\section{Systematic uncertainties}
\label{sec:systematics}

The sources of systematic uncertainties can be grouped into four categories:
\begin{itemize}
\item Uncertainties associated with the normalization of the background sources. These are treated as uncorrelated among backgrounds.
\item Uncertainties related to the modelling of instrumental effects in the simulation such as those arising from the object identification, reconstruction, and energy scales. These are typically correlated across all processes.
\item Uncertainties in the measurement of the integrated luminosity collected by CMS that are also correlated among all processes estimated from simulation.
\item Theoretical and modelling uncertainties that affect only the \WZ process.
\end{itemize}
For those backgrounds that have a significant contribution to the SR (\ZZ, \ttX, and \Xg), no additional normalization uncertainty is included, as it is treated as an unconstrained parameter in the fit. As the remaining irreducible backgrounds are small and the uncertainty in their estimation contributes negligibly to the total uncertainty of the measurement, we use the experimentally measured uncertainties on each to reduce theoretical contributions to the final uncertainty: 15\% for \tZq~\cite{CMS:2018sgc}, 50\% for \VVV~\cite{CMS:2020hjs}, and 25\% for \VH~\cite{CMS:2018nsn}. Finally, a normalization uncertainty of 30\% is assigned for the non-prompt lepton background, as detailed in Section~\ref{sec:selection}.

The uncertainties related to the mismodelling of instrumental effects in the simulation are estimated by varying the relevant parameters within their uncertainties and recomputing the event yields. The following sources of systematic uncertainties are considered: lepton identification and isolation efficiencies, jet energy scale and resolution, \PQb tagging efficiency, \ptmiss scale and resolution, and pileup modelling.

The lepton identification and isolation efficiencies are estimated using the tag-and-probe technique~\cite{CMS-PAS-EWK-10-002} and are affected by the size of the data set in $\PZ\to\Pellp\Pellm$ enriched regions, where the efficiencies are derived. Both efficiencies and their uncertainties are estimated separately for electrons and muons, with the average uncertainties of $\sim$1\% per electron and $\sim$0.5\% per muon. The statistical component of these uncertainties, corresponding to the size of the data set in which the tag-and-probe technique has been applied, is also included. Uncertainties in the energy scale of electrons introduce a smaller effect that induces variations in the lepton \pt. This effect is negligible for muons.

The efficiency of the trigger selection is measured in data selected using an independent set of \ptmiss triggers and only the requirement of three tight leptons passing the \pt requirements of the SR described in Section~\ref{sec:selection}. The total uncertainty in the trigger efficiency measurement is split into a systematic component for the estimated measurement bias in the selection of the data set, and an additional statistical component originating from the limited size of this data set. Each component is applied in a decorrelated manner to different parts of the entire data set, and the uncertainties amount range between 0.5-1.0\% and from 0.5-1.6\% for the systematic and statistical sources, respectively. Since the central estimation of the trigger efficiency is high (over 99\%), the Clopper--Pearson estimation~\cite{Clopper:1934sws} of the associated confidence interval leads to slightly asymmetric effects that slightly increase the uncertainty on the downward variations.

The energy scale and resolution of the reconstructed jets has an associated uncertainty, which is propagated to the \ptmiss, causing a small effect in the total cross section. The \PQb tag veto requirement is affected by the uncertainties in the efficiency of the \PQb tagging procedure and the mistagging rate. The effect of these uncertainties is minimized by choosing a tight working point of the \PQb tagging algorithm. They are split into the uncertainties in the tagging of heavy-flavour, and light-flavour and gluon jets as two decorrelated sources.

The uncertainty assigned to the number of pileup events in simulation is obtained by modifying the inelastic \pp cross section up and down by 4.6\%~\cite{CMS:2018mlc} around its nominal value and propagating the effect to all simulated samples. The measurement of the integrated luminosity recorded by CMS introduces an additional uncertainty of 1.4\%~\cite{CMS-PAS-LUM-22-001, CMS:LUM-17-003} into the analysis. Uncertainties due to the variations of the PDFs and the missing higher order corrections in QCD, are also included in the fits. Effects due to the choice of \muF and \muR are evaluated by varying the parameters up and down independently by a factor of two around the nominal \POWHEG $\mu_0$ value under the constraint $0.5\leq\muF/\muR\leq2$. The envelope of these six variations is taken as the estimated uncertainty for the missing higher order QCD corrections. Parametric (PDF+\alpS) uncertainties are estimated using the NNPDF3.1 set.  Uncertainties related to the modelling of the parton shower for the \WZ signal have been estimated separately for the initial- and final-state radiation (ISR and FSR) contributions by varying the associated energy scales in simulation.
Finally the uncertainty coming from the finite size of the MC samples is estimated using the Barlow--Beeston method~\cite{Barlow:1993dm}. A summary of all the input uncertainties considered in the analysis is shown in Table~\ref{tab:systsCorr}.

\begin{table}[!ht]
\centering
\topcaption{Summary of the input relative uncertainties. Numbers are presented in percentages over the predicted signal yields of the associated process on which they have an effect. All uncertainties are treated as shape variations on the templates used for the fit, with the exception of the normalization uncertainties in the backgrounds that are treated as flat variations of the corresponding yield.}
\label{tab:systsCorr}
\renewcommand{\arraystretch}{1.1}
\begin{tabular}{lcc}
    Source                          & Uncertainty (\%) & Processes \\
    \hline
    Integrated luminosity           &   1.4       &   All MC \\
    Trigger efficiencies            &   0.5--1.6   &   All MC \\
    \PQb tagging (heavy)            &   0.1       &   All MC \\
    \PQb tagging (light/gluon)      &   $<$0.1    &   All MC \\
    Pileup                          &   0.4--1.7       &   All MC \\
    Jet energy scale                &   1.1       &   All MC \\
    Electron reconstruction         &   3.8       &   All MC \\
    Electron ident. efficiency      &   3.5       &   All MC \\
    Electron energy scale           &   0.3       &   All MC \\
    Muon efficiencies               &   1.1       &   All MC \\[\cmsTabSkip]
    Non-prompt bkg. normalization   &   30        &   Non-prompt \\
    \VVV normalization              &   50        &   \VVV \\
    \tZq normalization              &   15        &   \tZq \\
    \VH normalization               &   25        &   \VH  \\[\cmsTabSkip]
    ISR                             &   0.1       &   \WZ  \\
    FSR                             &   0.8       &   \WZ \\
    \WZ theory (\muR, \muF, PDF)    &   0.6       &   \WZ \\
\end{tabular}
\end{table}

\section{Fit strategy and results}
\label{sec:results}

The \WZ production cross section is measured in the detector-level phase space defined as SR in Section~\ref{sec:selection} in each of the four flavour-exclusive channels (\eee, \eem, \emm, and \mmm), and in the flavour-inclusive channel. This is motivated by the fact that final states with three leptons of the same-flavour (\eee and \mmm) include additional contributions to the production modes due to interference contributions from diagrams exchanging same flavour final-state leptons. The additional combinatorics of the two same-charged leptons result in slightly larger cross sections. At the same time, further splitting the final states into different lepton flavour compositions leads to both the possibility of analysing deviations in the lepton flavour universality in the \PW and \PZ boson decays, and of studying possible experimental biases in the determination of electron and muon efficiencies.

The observed \WZ cross section at the detector-level is extrapolated to two different particle-level phase spaces, refered to as ``fiducial'' (FR) and ``total'' (TR) regions, respectively. The TR is designed to provide a more inclusive measurement of the \WZ cross section, with fewer requirements on the properties of the process. In contrast, the FR is motivated by the aim to reduce as comprehensively as possible the detector acceptance effects on the theoretical uncertainties that affect the measurement. As a result, the FR closely mirrors the SR selection criteria outlined in Section~\ref{sec:selection}, including lepton kinematic requirements and the exclusion of $\WZ\to\PGt\PGt$. Both these regions are defined after hadronization and parton shower using generator-level quantities. Leptons at this level are ``dressed'' by adding the transverse momenta of generator-level photons within a cone of $\DR(\Pell,\PGg)<0.1$ to their own momenta. In addition, dressed leptons are associated to the \PW and \PZ bosons using the same tagging algorithm described in Section~\ref{sec:selection}. The proportion of non-FR events in the \WZ sample that pass detector-level requirements is approximately 5\% in all the different decay channels as well as in the inclusive channel, and it is mainly attributed to leptonic decays of at least one of the $\PGt$ lepton in any of the $\WZ\to\PGt+X$ decay modes, where X accounts for any lepton flavour. The proportion of FR events inside the TR, known as the acceptance, is about 17\% in the inclusive channel, or equivalently 4\% per decay channel. These numbers have been computed considering the \WZ\ \POWHEG signal sample, and cross-checked with an alternative \WZ \MGvATNLO sample. Nevertheless, they are not directly used as an input of the analysis. Instead, the non-FR events are estimated by treating them as background contributions in the extraction of the fiducial results, and as part of the signal in the TR results. The normalization of the WZ signal contributions in both cases is then measured directly in the maximum likelihood fit detailed below. The requirements of the FR and TR selections are summarised in Table~\ref{tab:fiducial_total_requirements}.

\begin{table}[ht!]
\centering
\topcaption{Definition of the fiducial and total regions. The lepton related quantities refer to ``dressed'' leptons. A dash symbol is used in cases the selection is not considered for the corresponding phase space.}
\label{tab:fiducial_total_requirements}
\renewcommand{\arraystretch}{1.1}
\begin{tabular}{ccc}
    Region & Fiducial & Total \\\hline
    Lepton definition & Dressed (\Pe, \PGm) & Dressed (\Pe, \PGm, \PGt) \\
    $\Nell=3$ & \checkmark & \checkmark \\
    $\pt\{\lepZI,\lepZII,\lepW\}>\{25,15,25\}\GeV$ & \checkmark & \NA \\
    $\abseta\{\lepZI,\lepZII,\lepW\}<\{2.5,2.5,2.5\}$ & \checkmark & \NA \\
    $\NOSSF=1$ & \checkmark & \checkmark \\
    $60<\mllmZ<120\GeV$ & \checkmark & \checkmark \\
    $\minmll>4\GeV$ & \checkmark & \checkmark \\
    $\mlll>100\GeV$ & \checkmark & \NA \\
\end{tabular}
\end{table}

Results at the fiducial- and total-level are obtained from two separate maximum likelihood fits to the distributions of the number of events in different flavour channels considering the SR and all the CRs described in Section~\ref{sec:background}, using the \textsc{combine} tool~\cite{CMS:2024onh}.  In the fits, the normalization of the signal contribution to the \WZ process and the \ZZ, \ttZ, and \Xg backgrounds are considered unconstrained parameters in the fit. The systematic uncertainties described in Section~\ref{sec:systematics} are included as nuisance parameters with gaussian constraints correlated across SR and CRs for all shape variations, and as log-normal functions for those sources that only have a normalization effect. In the following, all presented results consider the fit results using the TR definitions unless stated otherwise. In both fits the shape for the variations originating from theoretical sources of uncertainty are normalized in a way that ensures that the sample cross section of the \WZ process is the same in both the nominal templates and all the variations. This ensures that normalization effects on the theoretical uncertainties of the \WZ variations do not have an effect in the measurement, keeping only the contribution that changes the overall acceptance of the FR and TR selections. The same procedure for the extraction of the cross section results is used as well for the flavour-exclusive measurements.

\begin{figure}[!ph]
\centering
\includegraphics[width=0.48\textwidth]{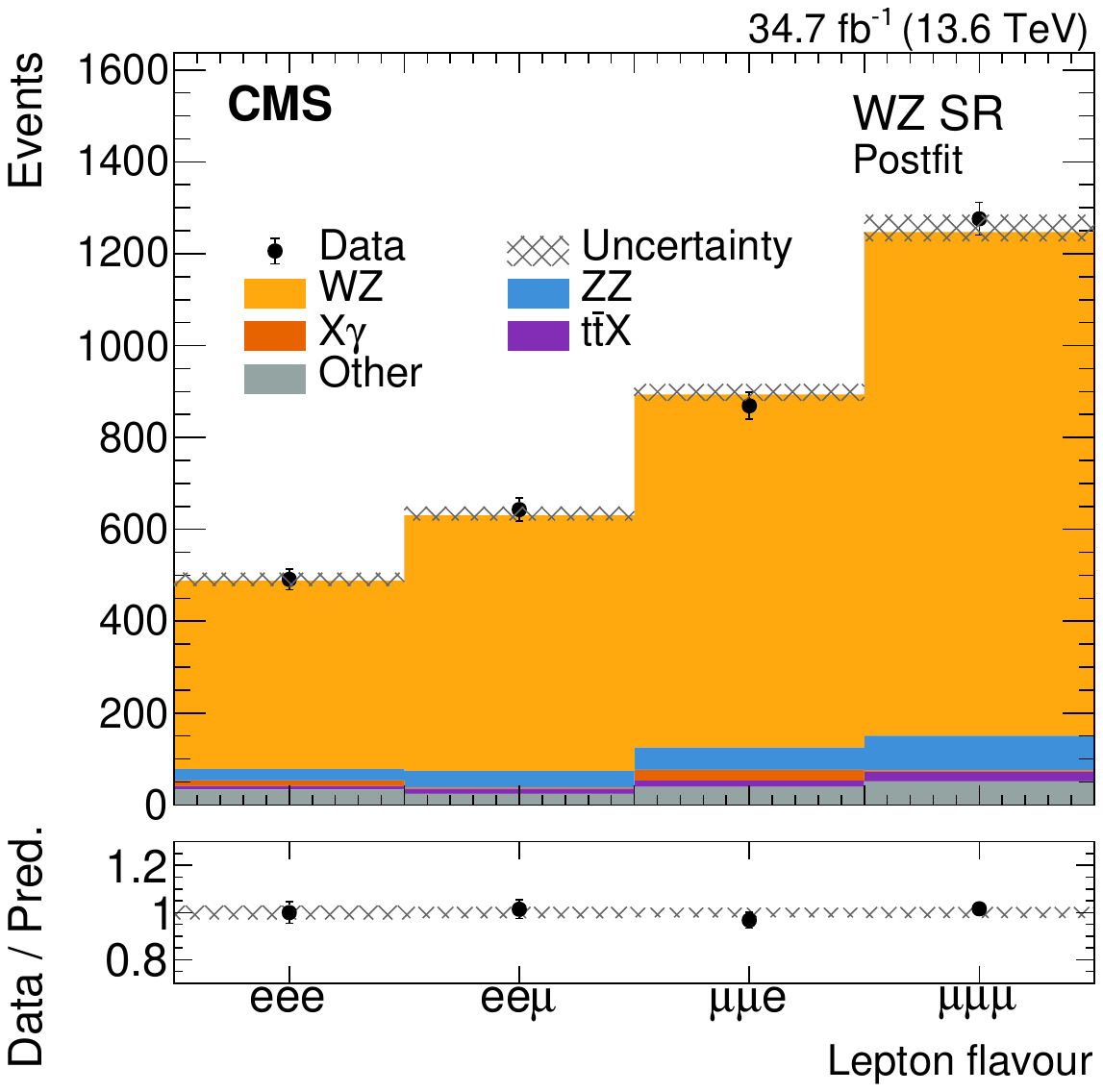}%
\hfill%
\includegraphics[width=0.48\textwidth]{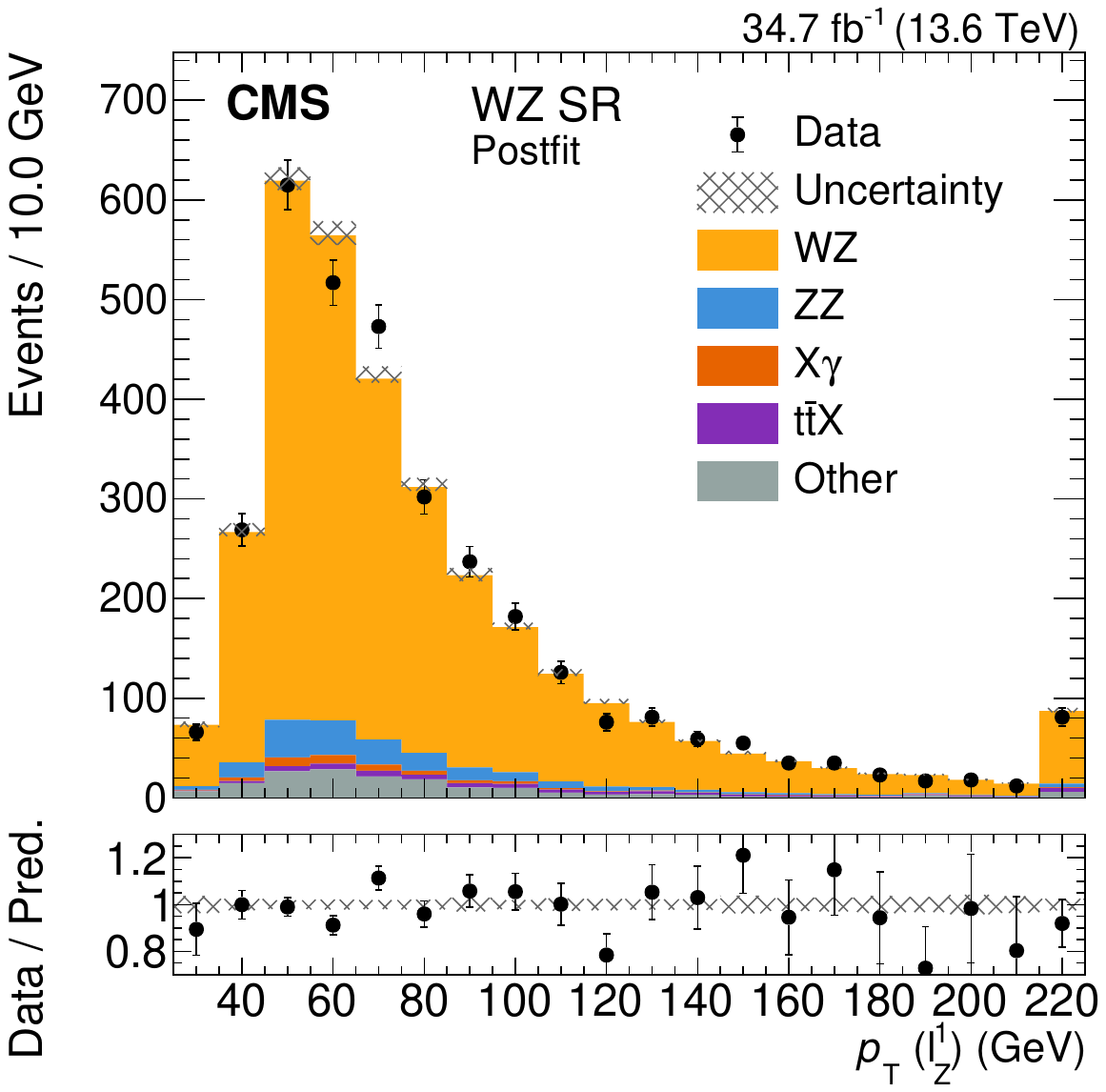} \\[1ex]
\includegraphics[width=0.48\textwidth]{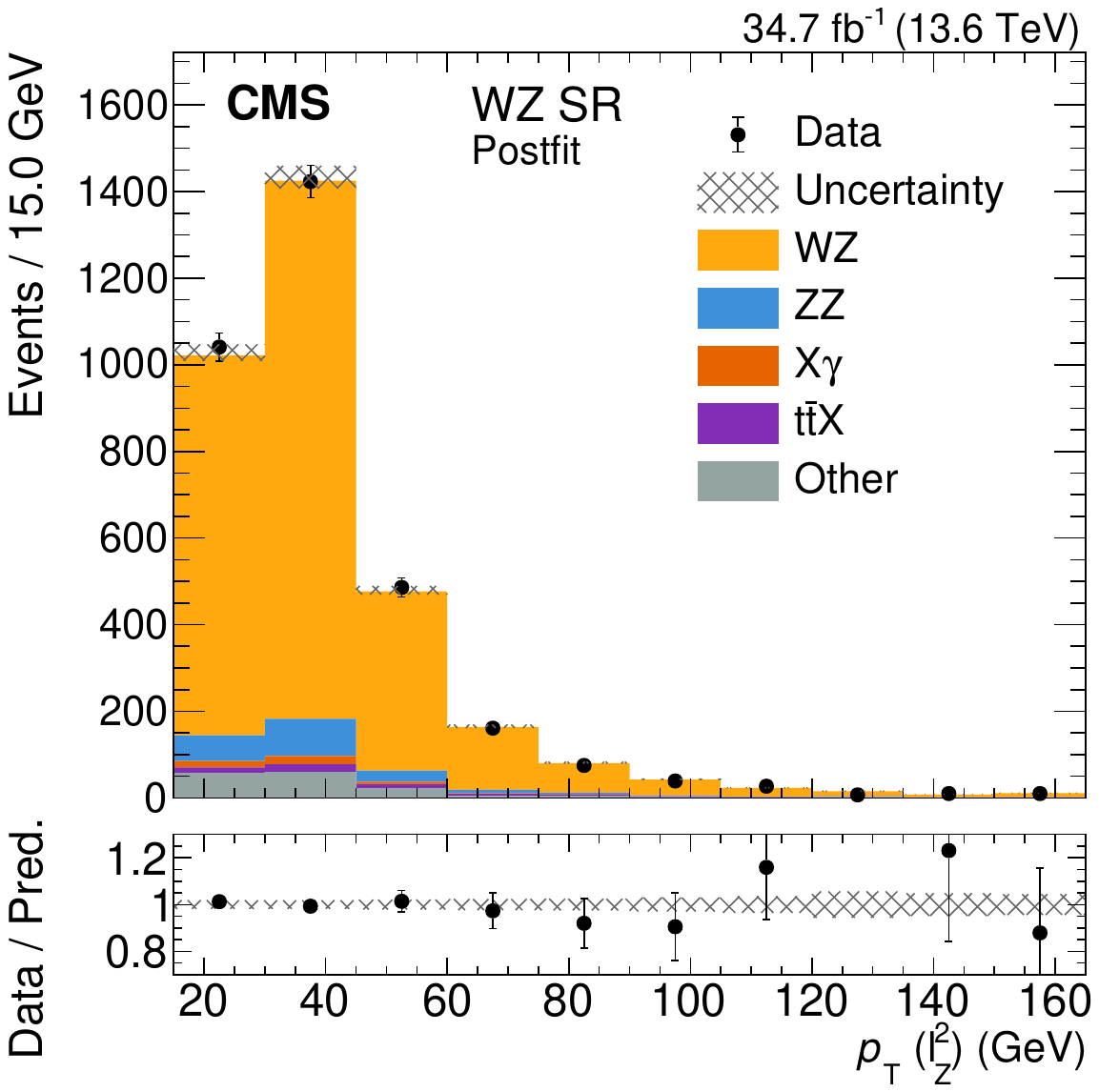}%
\hfill%
\includegraphics[width=0.48\textwidth]{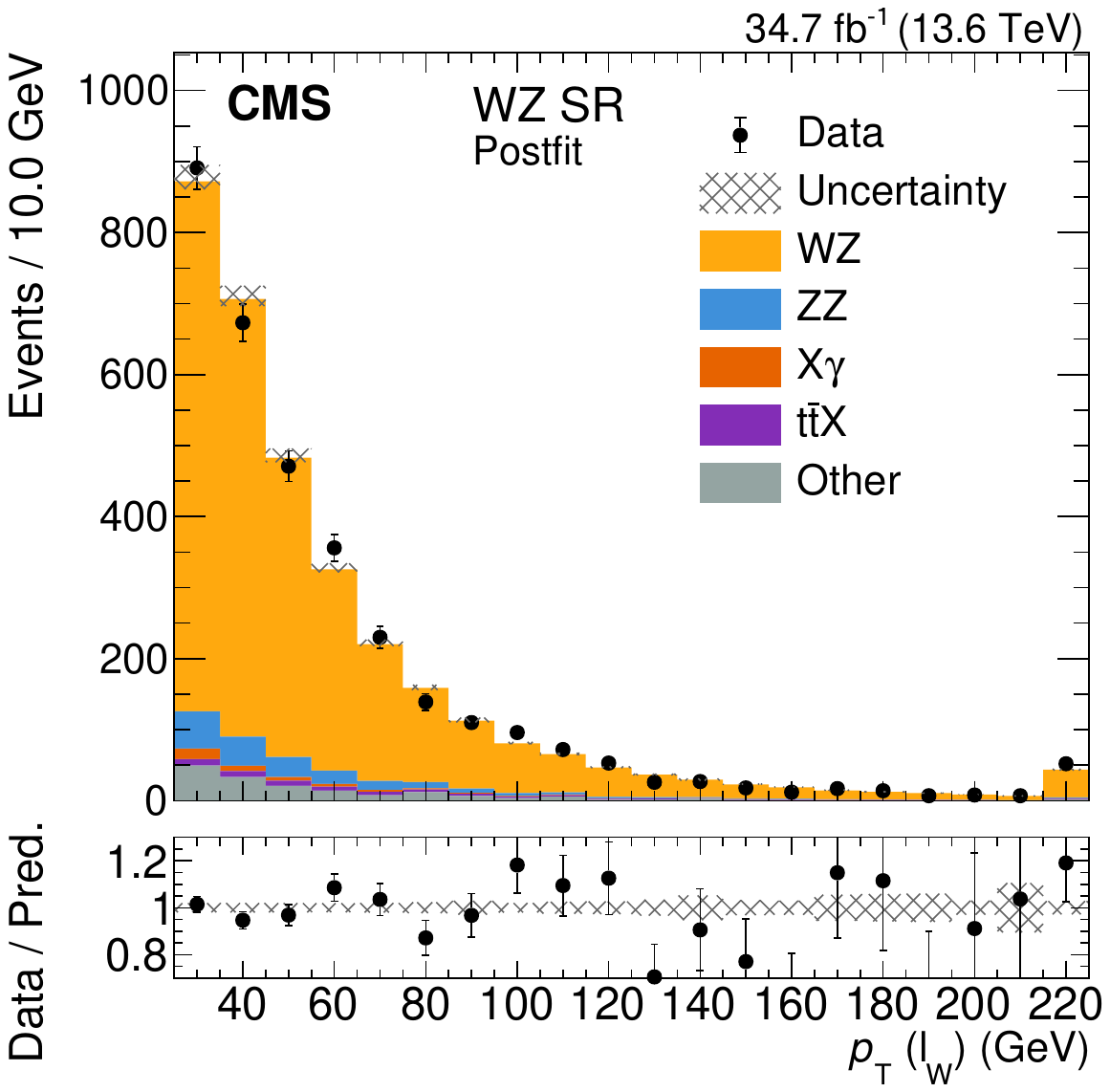}
\caption{Distributions of several observables in the SR accounting for the fit to data. From upper left (flavour composition), upper right (\pt of \lepZI), lower left (\pt of \lepZII), and lower right (\pt of \lepW). The hatched band includes all systematic uncertainties in the MC prediction. The vertical bars of the data account for the statistical uncertainty. When present, overflow events are included in the last bin of the observables. The ratio panels show the ratio between data (black markers) with respect to the total prediction after the fit to data. Processes with a small contribution to this region are grouped in the ``Other'' category.}
\label{fig:lepKins}
\end{figure}

\begin{figure}[!ph]
\centering
\includegraphics[width=0.48\textwidth]{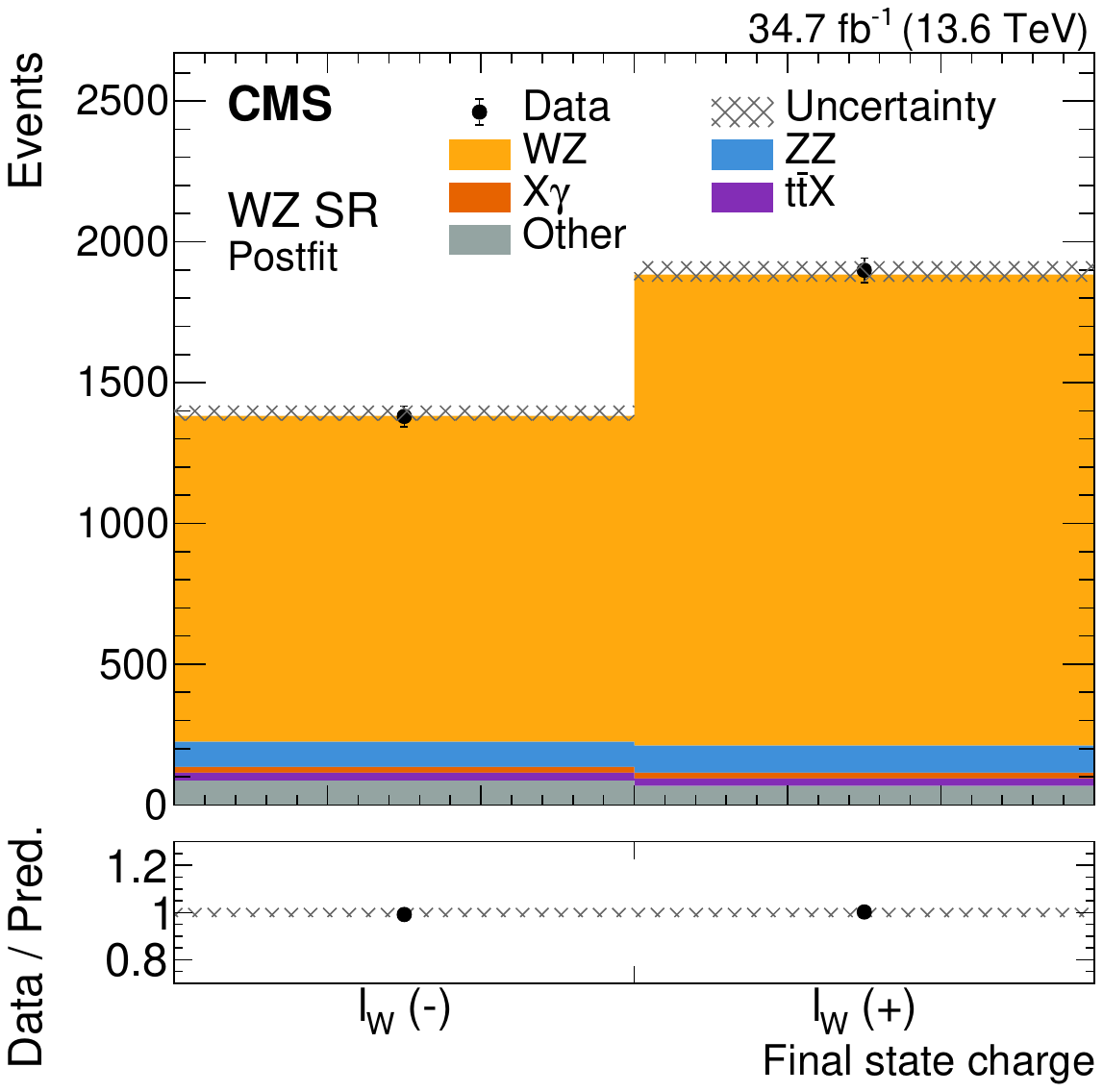}%
\hfill%
\includegraphics[width=0.48\textwidth]{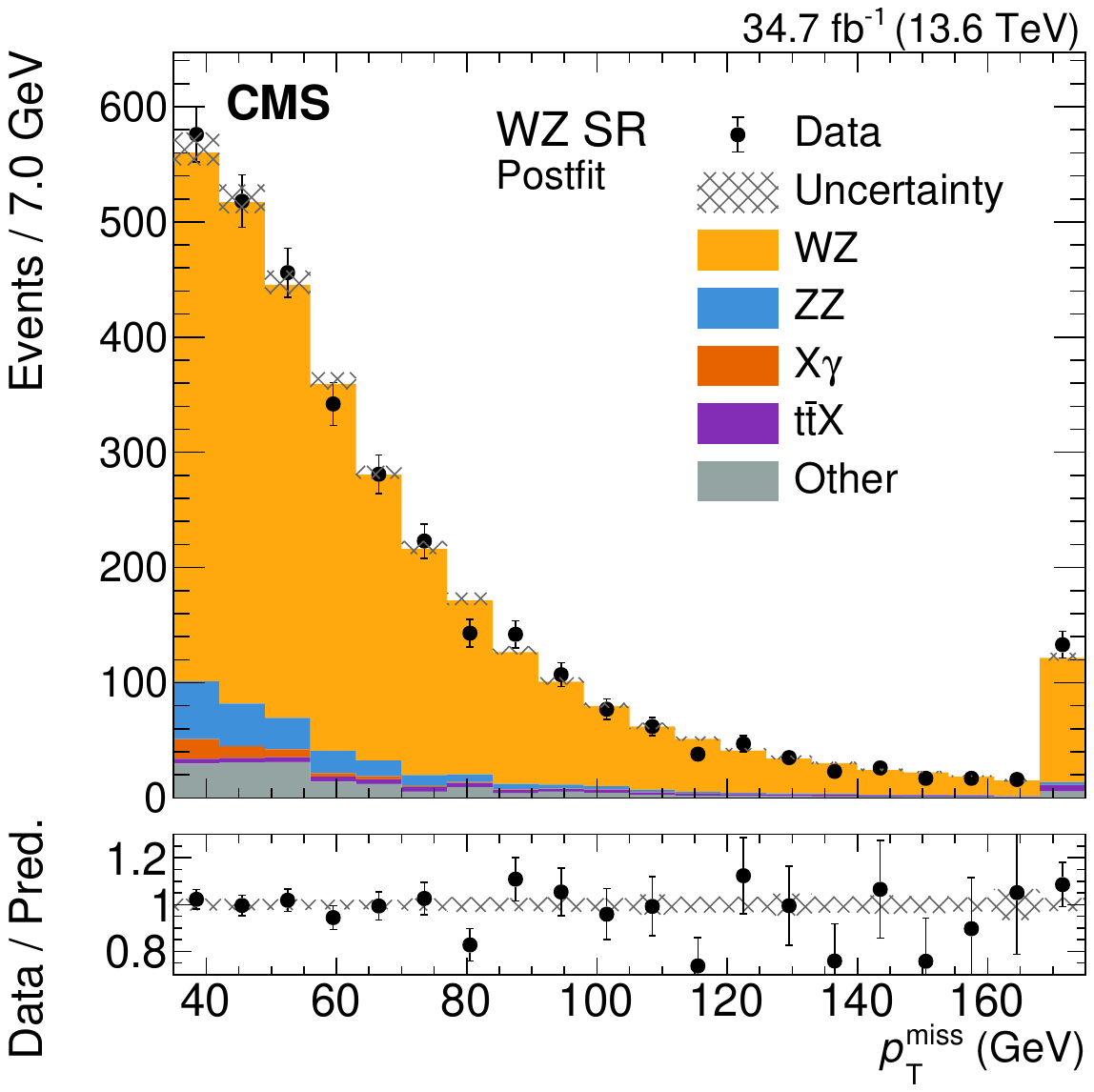}\\[1ex]
\includegraphics[width=0.48\textwidth]{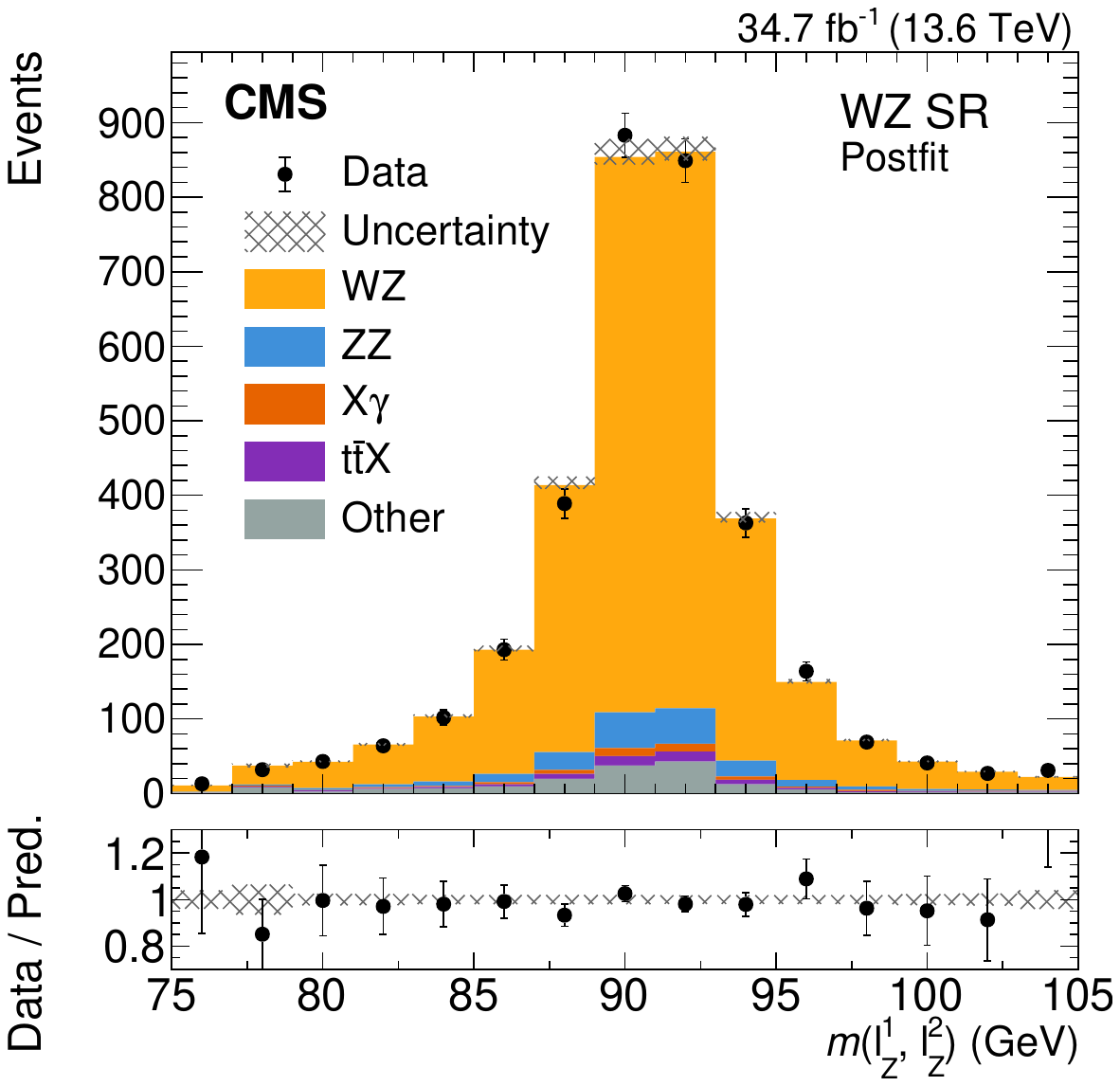}%
\hfill%
\includegraphics[width=0.48\textwidth]{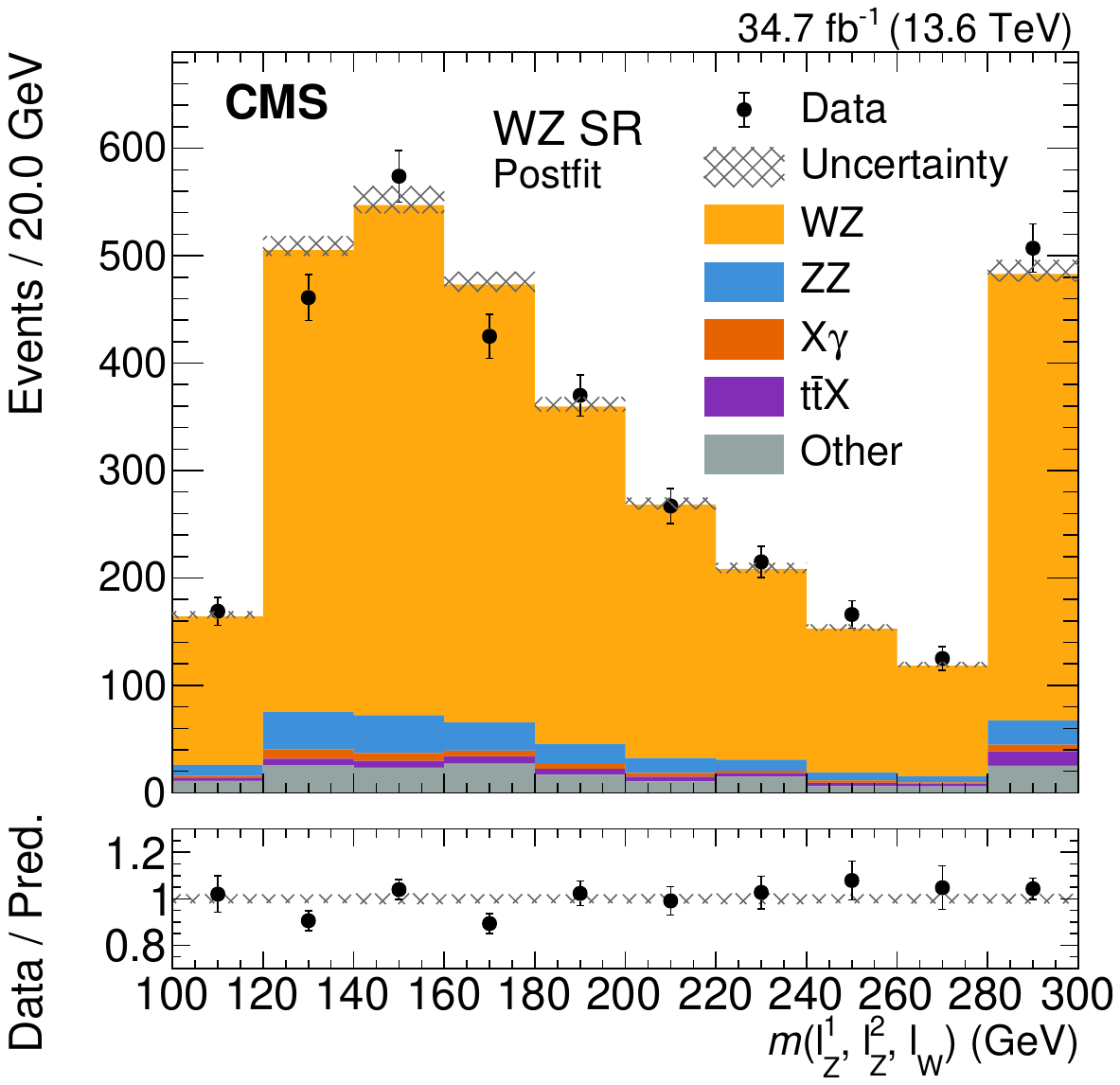}
\caption{Distributions of several observables in the SR accounting for the fit to data. From upper left (sum of charge of the final-state leptons), upper right (missing transverse momentum), lower left (invariant mass of the two leptons assigned to the \PZ boson decay), and lower right (invariant mass of the trilepton system). The hatched band includes all systematic uncertainties. When present, overflow events are included in the last bin of the observables. The ratio panels show the ratio between data (black markers) with respect to the total prediction after the fit to data. Processes with a small contribution to this region are grouped in the ``Other'' category.}
\label{fig:eventkin}
\end{figure}

Figures~\ref{fig:lepKins} and~\ref{fig:eventkin} show several relevant distributions in the SR accounting for the fit to data, including the distribution used in the fit as well as other kinematic properties of the selected leptons. Observables that do not enter the fit are obtained by propagating the fit results to the observables before the fit. The event yields per process and flavour channel after the fit are shown in Table~\ref{tab:yields}. Good agreement after the fit is observed in all the observables.

\begin{table}[!ht]
\centering
\topcaption{Number of selected events and their total uncertianty (by flavour channel) for the relevant processes in the signal region of the analysis accounting for the fit to data. The ``Background'' yield is the sum of all processes that are not \WZ signal, and ``Prediction'' is the sum of all the processes, including \WZ.}
\label{tab:yields}
\renewcommand{\arraystretch}{1.1}
\begin{tabular}{lr@{${}\pm{}$}lr@{${}\pm{}$}lr@{${}\pm{}$}lr@{${}\pm{}$}lr@{${}\pm{}$}l}
    Process & \multicolumn{2}{c}{\eee} & \multicolumn{2}{c}{\eem} & \multicolumn{2}{c}{\mme} & \multicolumn{2}{c}{\mmm} & \multicolumn{2}{c}{Inclusive} \\
    \hline
    Non-prompt &    25 &   7  &    13 &   5  &    24 &   7  &    30 &  10  &    93 &  15\\
    \ZZ &    25 &   2  &    37 &   1  &    49 &   3  &    75 &   3  &   186 &   5\\
    \Xg &    12 &   2  &   2.5 & 0.3  &    24 &   2  &   3.2 & 0.5  &    41 &   3\\
    \ttX &   8.0 & 0.8  &    11 &   1  &    14 &   1  &    21 &   2  &    54 &   3\\
    \VVV &     4 &   1  &     5 &   2  &     7 &   3  &    10 &   4  &    27 &   5\\
    \VH &   3.0 & 0.5  &   3.8 & 0.7  &     5 &   1  &     9 &   2  &    20 &   2\\
    \tZq &   4.2 & 0.5  &   5.3 & 0.6  &   7.5 & 0.9  &    11 &   1  &    28 &   2\\[\cmsTabSkip]
    Background &    82 &   8  &    78 &   5  &   130 &   9  &   160 &  11  &   450 &  17\\
    \WZ &   410 &  10  &   556 &  12  &   768 &  14  &   1096 &  22  &   2830 &  31\\
    Prediction &   491 &  13  &   634 &  13  &   898 &  16  &   1256 &  24  &   3280 &  34\\[\cmsTabSkip]
    Data & \multicolumn{2}{c}{491} & \multicolumn{2}{c}{643} & \multicolumn{2}{c}{869} & \multicolumn{2}{c}{1276} & \multicolumn{2}{c}{3279} \\
\end{tabular}
\end{table}

\begin{figure}[!ph]
\centering
\includegraphics[width=0.9\textwidth]{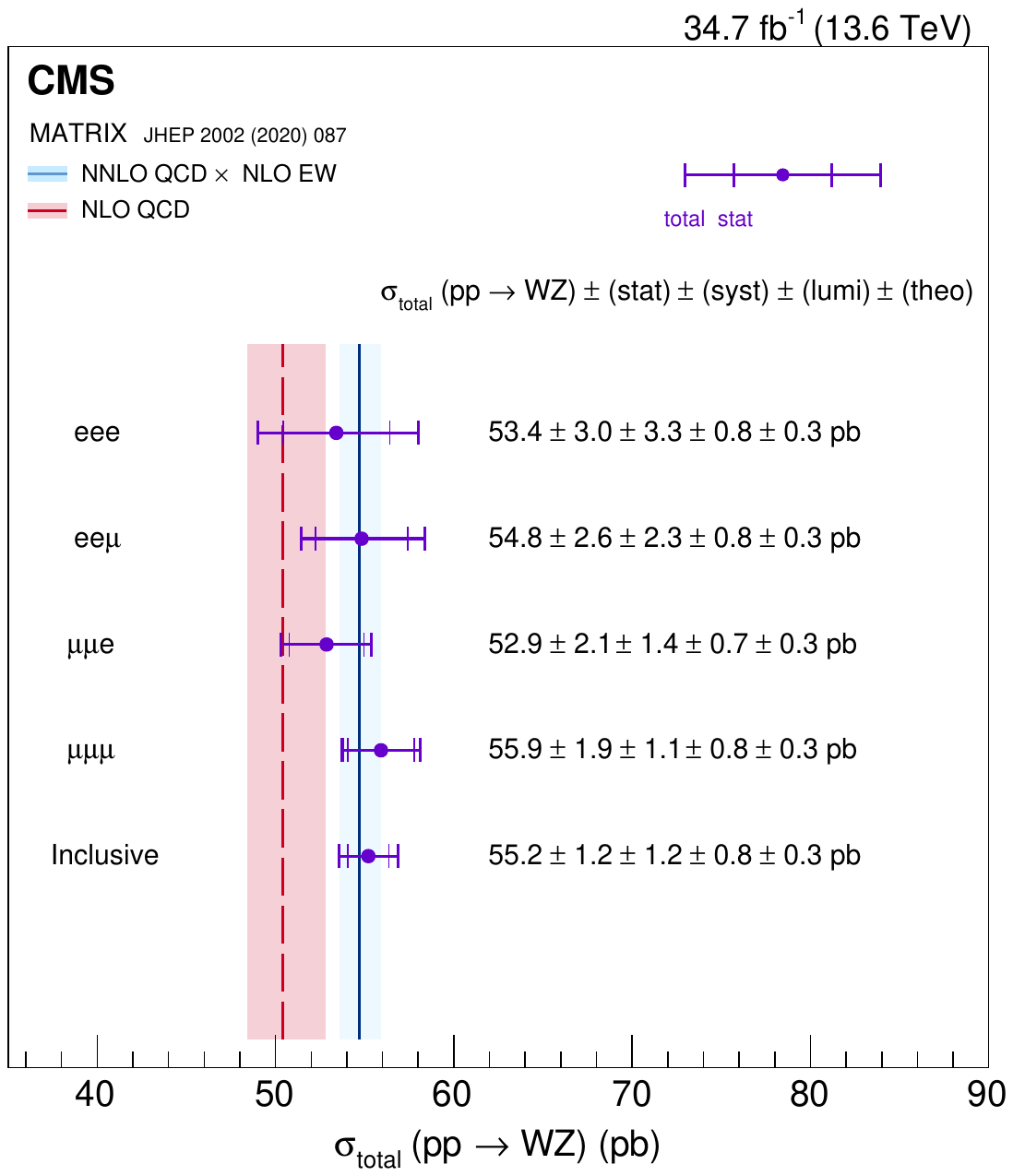}
\caption{Total \WZ production cross section for each of the flavour-exclusive and for the flavour-inclusive categories. The vertical bands show different theoretical predictions for the \WZ cross section at NLO in QCD (red dashed line) and \NNLOQCDNLOEW (blue solid line), as well as their corresponding scale uncertainties. For each measurement, the best fit value is denoted with a purple point, with two delimiters on the error bars that account for the statistical and total uncertainties.}
\label{fig:inc}
\end{figure}

\begin{figure}[!th]
\centering
\includegraphics[width=0.9\textwidth]{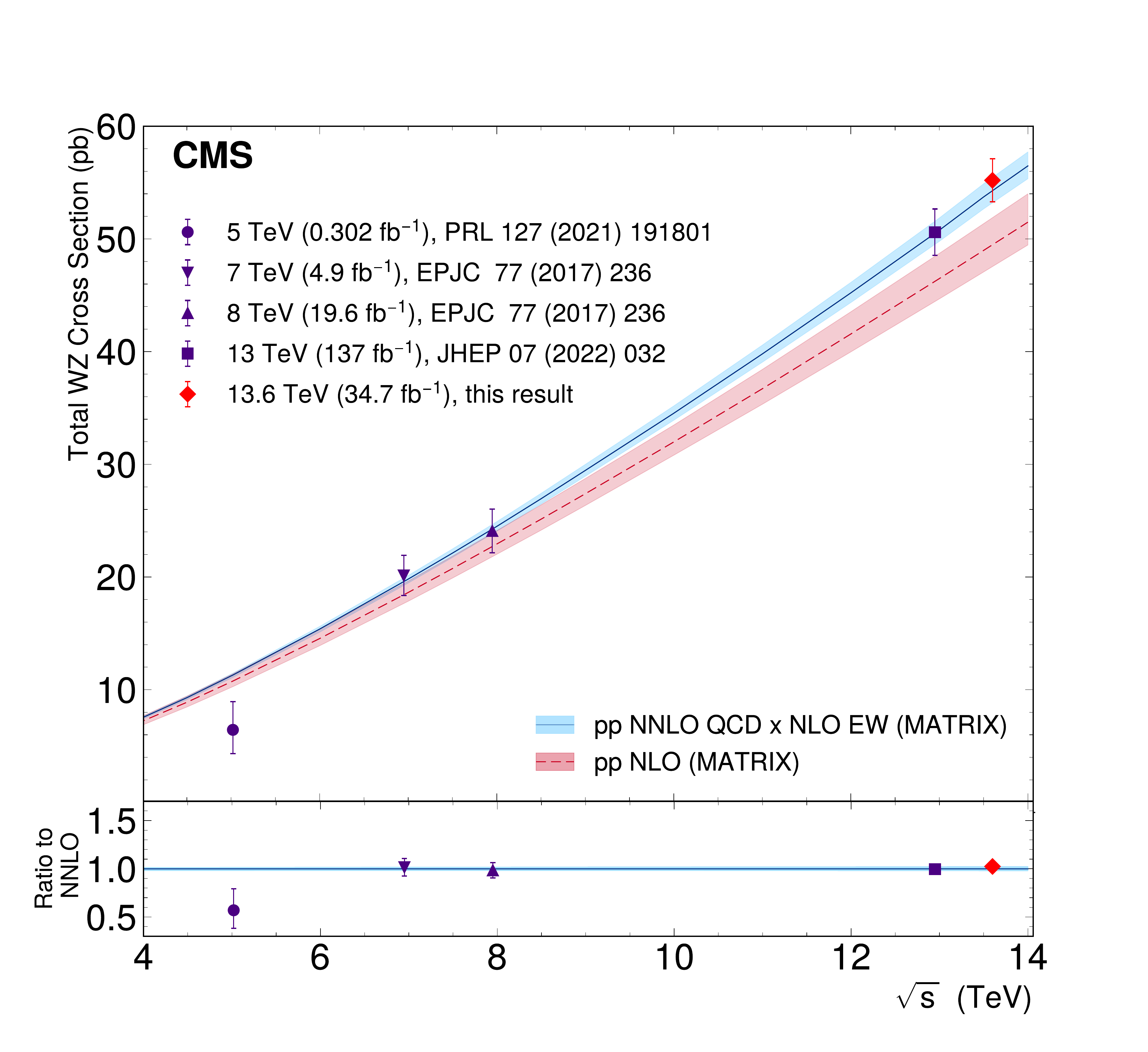}
\caption{Measurement obtained in this analysis (red filled marker) together with other \WZ production cross section measurements at different centre-of-mass energies by the CMS~\cite{CMS:2021pqj, CMS:2016qth, CMS:2021icx} Collaboration, compared to the \NNLOQCDNLOEW predictions, as well as the NLO QCD prediction, computed in all cases with \MATRIX.}
\label{fig:sigmaSumm}
\end{figure}

The measured fiducial cross sections are reported in Table~\ref{tab:fiducial} for the different flavour channels, including a comparison with theoretical predictions computed from the signal \POWHEG sample, as well as the next-to-NLO (NNLO) QCD and \NNLOQCDNLOEW prediction from \MATRIX~\cite{Grazzini:2019jkl}. In all \MATRIX predictions we have adopted the main setup for the SM parameters, scale and PDF choices, and have only modified the requirements of the leptons in order to mimic either the fiducial or total definitions. Similarly, results for the total cross section measurement and their comparison to different predictions at different orders of the QCD and EW expansion can be found in Table~\ref{tab:total} and Fig.~\ref{fig:inc}. The relative uncertainty in the measurement in the inclusive category of about 3.4\% is compatible between the fiducial-level and total-level results. A breakdown of the different sources of systematic uncertainty and their impact in the result can be found in Table~\ref{tab:table_syst}. The values of the normalization parameters in the combined fit corresponding to each of the individual backgrounds are: $r_{\ZZ} = 0.96\pm0.06$, $r_{\ttZ} = 1.18 \pm 0.15$, and $r_{\Xg}=1.10 \pm 0.13$. The measurements presented in this paper, as well as all previous \WZ measurements by CMS at different centre-of-mass energies, support the existence of large corrections beyond the \NLOQCD and the good compatibility with the latest \NNLOQCDNLOEW predictions, as summarised in Fig.~\ref{fig:sigmaSumm}.  

\begin{table}[!ht]
\centering
\topcaption{Measured fiducial cross sections and their corresponding uncertainties for the flavour-exclusive and flavour-inclusive categories.
The predictions from both \POWHEG at NLO in QCD and LO EW as well as several ones obtained from \MATRIX (\NLOQCD, \NNLOQCD, \NNLOQCDNLOEW) are also included.}
\label{tab:fiducial}
\renewcommand{\arraystretch}{1.2}
\cmsTable{\begin{tabular}{lcc}
    Category & Accuracy & Fiducial cross section (fb) \\ \hline
    \multirow{4}{*}{\eee}
    & \POWHEG, \NLOQCD       & $68.0\,^{+2.3}_{-2.1}\scale\pm1.0\PDF$ \\
    & \MATRIX, \NLOQCD       & $69.9\,^{+3.9}_{-3.1}\scale$ \\
    & \MATRIX, \NNLOQCD      & $77.0\,^{+1.8}_{-1.7}\scale$ \\
    & \MATRIX, \NNLOQCDNLOEW & $75.4\,^{+1.7}_{-1.6}\scale$ \\
    & Measured               & $72.0\pm4.0\stat\pm4.5\syst\pm1.0\lum\pm0.4\thy$ \\[\cmsTabSkip]
    \multirow{4}{*}{\eem}
    & \POWHEG, \NLOQCD       & $68.0\,^{+2.3}_{-2.1}\scale\pm1.0\PDF$ \\
    & \MATRIX, \NLOQCD       & $68.7\,^{+3.8}_{-3.0}\scale$ \\
    & \MATRIX, \NNLOQCD      & $75.0\,^{+1.8}_{-1.6}\scale$ \\
    & \MATRIX, \NNLOQCDNLOEW & $73.4\,^{+1.7}_{-1.5}\scale$ \\
    & Measured               & $73.9\pm3.5\stat\pm3.1\syst\pm1.1\lum\pm0.4\thy$ \\[\cmsTabSkip]
    \multirow{4}{*}{\mme}
    & \POWHEG, \NLOQCD       & $68.0\,^{+2.3}_{-2.1}\scale\pm1.0\PDF$ \\
    & \MATRIX, \NLOQCD       & $68.7\,^{+3.8}_{-3.0}\scale$ \\
    & \MATRIX, \NNLOQCD      & $75.0\,^{+1.8}_{-1.6}\scale$ \\
    & \MATRIX, \NNLOQCDNLOEW & $73.4\,^{+1.7}_{-1.5}\scale$ \\
    & Measured               & $71.2\pm2.9\stat\pm2.0\syst\pm1.0\lum\pm0.4\thy$ \\[\cmsTabSkip]
    \multirow{4}{*}{\mmm}
    & \POWHEG, \NLOQCD       & $68.0\,^{+2.3}_{-2.1}\scale\pm1.0\PDF$ \\
    & \MATRIX, \NLOQCD       & $69.9\,^{+3.9}_{-3.1}\scale$ \\
    & \MATRIX, \NNLOQCD      & $77.0\,^{+1.8}_{-1.7}\scale$ \\
    & \MATRIX, \NNLOQCDNLOEW & $75.4\,^{+1.7}_{-1.6}\scale$ \\
    & Measured               & $75.3\pm2.5\stat\pm1.5\syst\pm1.1\lum\pm0.4\thy$ \\[\cmsTabSkip]
    \multirow{4}{*}{Inclusive}
    & \POWHEG, \NLOQCD       & $271.9\,^{+9.0}_{-8.5}\scale\pm3.8\PDF$ \\
    & \MATRIX, \NLOQCD       & $277.1\,^{+15.3}_{-12.3}\scale$ \\
    & \MATRIX, \NNLOQCD      & $304.0\,^{+7.1}_{-6.6}\scale$ \\
    & \MATRIX, \NNLOQCDNLOEW & $297.7\,^{+6.8}_{-6.3}\scale$ \\
    & Measured               & $297.6\pm6.4\stat\pm6.4\syst\pm4.2\lum\pm1.5\thy$ \\[\cmsTabSkip]
\end{tabular}}
\end{table}

\begin{table}[!ph]
\centering
\topcaption{Flavour-inclusive total cross section result. The predictions from both \POWHEG at NLO in QCD and LO EW as well as several ones obtained from \MATRIX (\NLOQCD, \NNLOQCD, \NNLOQCDNLOEW) are also included.}
\label{tab:total}
\renewcommand{\arraystretch}{1.2}
\begin{tabular}{lc}
    Accuracy               & Total cross section (pb) \\ \hline
    \POWHEG, \NLOQCD       & $50.5\,^{+2.6}_{-2.1}\scale \pm 1.1\PDF$ \\
    \MATRIX, \NLOQCD       & $50.4\,^{+2.3}_{-2.0}\scale$ \\
    \MATRIX, \NNLOQCD      & $55.0\,^{+1.2}_{-1.1}\scale$ \\
    \MATRIX, \NNLOQCDNLOEW & $54.7\,^{+1.2}_{-1.1}\scale$ \\[\cmsTabSkip]
    Inclusive (Measured)   & $55.2\pm1.2\stat\pm1.2\syst\pm0.8\lum\pm0.3\thy$ \\
\end{tabular}
\end{table}

\begin{table}[!pth]
\centering
\topcaption{Breakdown of different sources of systematic uncertainties and their relative impact in each channel, as well as in the inclusive measurement; as a percentage of the total uncertainty. The dash symbol indicates that the specific uncertainty does not apply.}
\label{tab:table_syst}
\renewcommand{\arraystretch}{1.1}
\begin{tabular}{lccccc}
    Source & Inclusive (\%) & \eee (\%) & \eem (\%) & \mme (\%) & \mmm (\%)\\
    \hline
    Integrated luminosity & 1.5 & 1.5 & 1.4 & 1.4 & 1.5\\
    Trigger efficiencies & 0.5 & 1.0 & 1.0 & 1.0 & 0.7\\
    \PQb tagging & 0.1 & 0.1 & 0.1 & 0.1 & 0.1\\
    Pileup & 0.4 & 0.6 & 0.8 & 0.2 & 0.4\\
    Jet energy scales & 0.9 & 1.3 & 0.7 & 1.1 & 0.7\\
    Electron reconstruction & 1.2 & 4.0 & 2.9 & 1.1 & \NA\\
    Electron ident. efficiencies & 0.7 & 3.6 & 2.4 & 1.1 & \NA\\
    Electron energy scale & 0.1 & 0.1 & 0.1 & 0.0 & \NA\\
    Muon efficiencies & 0.7 & \NA & 0.3 & 0.8 & 1.2\\[\cmsTabSkip]
    Non-prompt bkg. normalization & 0.7 & 1.6 & 0.5 & 0.7 & 0.7\\
    \VVV normalization & 0.4 & 0.4 & 0.4 & 0.4 & 0.4\\
    \tZq normalization & 0.1 & 0.1 & 0.1 & 0.1 & 0.1\\
    \ZZ normalization & 0.3 & 0.8 & 0.7 & 0.5 & 0.5\\
    \ttZ normalization & 0.3 & 0.7 & 0.6 & 0.4 & 0.5\\
    \Xg normalization & 0.2 & 0.7 & 0.3 & 0.4 & 0.2\\
    \VH normalization & 0.2 & 0.2 & 0.2 & 0.1 & 0.2\\[\cmsTabSkip]
    ISR/FSR & 0.3 & 0.5 & 0.2 & 0.4 & 0.3\\
    \WZ theory (\muR, \muF, PDF) & 0.6 & 0.6 & 0.6 & 0.6 & 0.6\\
    MC statistical & 0.5 & 1.9 & 0.9 & 1.0 & 0.9\\
    Statistical & 2.0 & 5.3 & 4.6 & 3.8 & 3.3\\
    Total & 3.4 & 8.5 & 6.5 & 5.0 & 4.3\\
\end{tabular}
\end{table}

\section{Summary}
\label{sec:summary}

The inclusive \WZ production cross section is measured in proton-proton collisions for the first time at a centre-of-mass energy of 13.6\TeV, using data collected during 2022 with the CMS detector, corresponding to an integrated luminosity of 34.7\fbinv. The production cross sections in the total and fiducial phase spaces are measured in the inclusive case as well as in four combinations of final state flavour composition. The cross sections are measured in a fiducial phase space as $\sigma_{\text{fiducial}}(\pp\to\WZ)=297.6\pm6.4\stat\pm6.4\syst\pm4.2\lum\pm1.5\thy\unit{fb}$, and in a less restricted phase space as $\sigma_{\text{total}}(\pp\to\WZ)=55.2\pm1.2\stat\pm1.2\syst\pm0.8\lum\pm0.3\thy\unit{pb}$. All these measurements are shown to be in good agreement with the SM predictions at \NNLOQCDNLOEW.

\begin{acknowledgments}
We congratulate our colleagues in the CERN accelerator departments for the excellent performance of the LHC and thank the technical and administrative staffs at CERN and at other CMS institutes for their contributions to the success of the CMS effort. In addition, we gratefully acknowledge the computing centres and personnel of the Worldwide LHC Computing Grid and other centres for delivering so effectively the computing infrastructure essential to our analyses. Finally, we acknowledge the enduring support for the construction and operation of the LHC, the CMS detector, and the supporting computing infrastructure provided by the following funding agencies: SC (Armenia), BMBWF and FWF (Austria); FNRS and FWO (Belgium); CNPq, CAPES, FAPERJ, FAPERGS, and FAPESP (Brazil); MES and BNSF (Bulgaria); CERN; CAS, MoST, and NSFC (China); MINCIENCIAS (Colombia); MSES and CSF (Croatia); RIF (Cyprus); SENESCYT (Ecuador); ERC PRG, RVTT3 and MoER TK202 (Estonia); Academy of Finland, MEC, and HIP (Finland); CEA and CNRS/IN2P3 (France); SRNSF (Georgia); BMBF, DFG, and HGF (Germany); GSRI (Greece); NKFIH (Hungary); DAE and DST (India); IPM (Iran); SFI (Ireland); INFN (Italy); MSIP and NRF (Republic of Korea); MES (Latvia); LMTLT (Lithuania); MOE and UM (Malaysia); BUAP, CINVESTAV, CONACYT, LNS, SEP, and UASLP-FAI (Mexico); MOS (Montenegro); MBIE (New Zealand); PAEC (Pakistan); MES and NSC (Poland); FCT (Portugal); MESTD (Serbia); MCIN/AEI and PCTI (Spain); MOSTR (Sri Lanka); Swiss Funding Agencies (Switzerland); MST (Taipei); MHESI and NSTDA (Thailand); TUBITAK and TENMAK (Turkey); NASU (Ukraine); STFC (United Kingdom); DOE and NSF (USA).
    
\hyphenation{Rachada-pisek} Individuals have received support from the Marie-Curie programme and the European Research Council and Horizon 2020 Grant, contract Nos.\ 675440, 724704, 752730, 758316, 765710, 824093, 101115353, 101002207, and COST Action CA16108 (European Union); the Leventis Foundation; the Alfred P.\ Sloan Foundation; the Alexander von Humboldt Foundation; the Science Committee, project no. 22rl-037 (Armenia); the Belgian Federal Science Policy Office; the Fonds pour la Formation \`a la Recherche dans l'Industrie et dans l'Agriculture (FRIA-Belgium); the F.R.S.-FNRS and FWO (Belgium) under the ``Excellence of Science -- EOS" -- be.h project n.\ 30820817; the Beijing Municipal Science \& Technology Commission, No. Z191100007219010 and Fundamental Research Funds for the Central Universities (China); the Ministry of Education, Youth and Sports (MEYS) of the Czech Republic; the Shota Rustaveli National Science Foundation, grant FR-22-985 (Georgia); the Deutsche Forschungsgemeinschaft (DFG), among others, under Germany's Excellence Strategy -- EXC 2121 ``Quantum Universe" -- 390833306, and under project number 400140256 - GRK2497; the Hellenic Foundation for Research and Innovation (HFRI), Project Number 2288 (Greece); the Hungarian Academy of Sciences, the New National Excellence Program - \'UNKP, the NKFIH research grants K 131991, K 133046, K 138136, K 143460, K 143477, K 146913, K 146914, K 147048, 2020-2.2.1-ED-2021-00181, TKP2021-NKTA-64, and 2021-4.1.2-NEMZ\_KI (Hungary); the Council of Science and Industrial Research, India; ICSC -- National Research Centre for High Performance Computing, Big Data and Quantum Computing and FAIR -- Future Artificial Intelligence Research, funded by the NextGenerationEU program (Italy); the Latvian Council of Science; the Ministry of Education and Science, project no. 2022/WK/14, and the National Science Center, contracts Opus 2021/41/B/ST2/01369 and 2021/43/B/ST2/01552 (Poland); the Funda\c{c}\~ao para a Ci\^encia e a Tecnologia, grant CEECIND/01334/2018 (Portugal); the National Priorities Research Program by Qatar National Research Fund; MCIN/AEI/10.13039/501100011033, ERDF ``a way of making Europe", and the Programa Estatal de Fomento de la Investigaci{\'o}n Cient{\'i}fica y T{\'e}cnica de Excelencia Mar\'{\i}a de Maeztu, grant MDM-2017-0765 and Programa Severo Ochoa del Principado de Asturias (Spain); the Chulalongkorn Academic into Its 2nd Century Project Advancement Project, and the National Science, Research and Innovation Fund via the Program Management Unit for Human Resources \& Institutional Development, Research and Innovation, grant B39G670016 (Thailand); the Kavli Foundation; the Nvidia Corporation; the SuperMicro Corporation; the Welch Foundation, contract C-1845; and the Weston Havens Foundation (USA).
\end{acknowledgments}

\bibliography{auto_generated}

\providecommand{\href}[2]{#2}\begingroup\raggedright\begin{thebibliography}{10}%
\makeatletter
\providecommand{\hrefCMSnoop }[0]{\@secondoftwo}%
\makeatother
\providecommand{\doi}{\texttt{doi:}\begingroup \urlstyle{tt}\Url}

\bibitem{Degrande:2012wf}
C.~Degrande\hrefCMSnoop {}{ { et~al.}, ``Effective field theory: a modern
  approach to anomalous couplings'',} \textit{ Annals Phys.} \textbf{ 335}
  (2013) 21,
  \href{http://dx.doi.org/10.1016/j.aop.2013.04.016}{\doi{10.1016/j.aop.2013.04.016}},
  \href{http://www.arXiv.org/abs/1205.4231}{\texttt{arXiv:1205.4231}}.

\bibitem{CMS:2016qth}
\hrefCMSnoop {}{{CMS Collaboration}, ``Measurement of the ${\PW\PZ}$ production
  cross section in ${\Pp\Pp}$ collisions at $\sqrt{s}=7$ and {8\TeV} and search
  for anomalous triple gauge couplings at $\sqrt{s}={8\TeV}$'',} \textit{ Eur.
  Phys. J. C} \textbf{ 77} (2017) 236,
  \href{http://dx.doi.org/10.1140/epjc/s10052-017-4730-z}{\doi{10.1140/epjc/s10052-017-4730-z}},
  \href{http://www.arXiv.org/abs/1609.05721}{\texttt{arXiv:1609.05721}}.

\bibitem{ATLAS:2016bkj}
\hrefCMSnoop {}{{ATLAS Collaboration}, ``Measurements of ${\PWpm\PZ}$
  production cross sections in ${\Pp\Pp}$ collisions at $\sqrt{s}={8\TeV}$ with
  the {ATLAS} detector and limits on anomalous gauge boson self-couplings'',}
  \textit{ Phys. Rev. D} \textbf{ 93} (2016) 092004,
  \href{http://dx.doi.org/10.1103/PhysRevD.93.092004}{\doi{10.1103/PhysRevD.93.092004}},
  \href{http://www.arXiv.org/abs/1603.02151}{\texttt{arXiv:1603.02151}}.

\bibitem{Khachatryan:2016tgp}
\hrefCMSnoop {}{{CMS Collaboration}, ``Measurement of the ${\PW\PZ}$ production
  cross section in ${\Pp\Pp}$ collisions at $\sqrt{s}={13\TeV}$'',} \textit{
  Phys. Lett. B} \textbf{ 766} (2017) 268,
  \href{http://dx.doi.org/10.1016/j.physletb.2017.01.011}{\doi{10.1016/j.physletb.2017.01.011}},
  \href{http://www.arXiv.org/abs/1607.06943}{\texttt{arXiv:1607.06943}}.

\bibitem{Aaboud:2016yus}
\hrefCMSnoop {}{{ATLAS Collaboration}, ``Measurement of the ${\PWpm\PZ}$ boson
  pair-production cross section in ${\Pp\Pp}$ collisions at $\sqrt{s}={13\TeV}$
  with the {ATLAS} detector'',} \textit{ Phys. Lett. B} \textbf{ 762} (2016) 1,
  \href{http://dx.doi.org/10.1016/j.physletb.2016.08.052}{\doi{10.1016/j.physletb.2016.08.052}},
  \href{http://www.arXiv.org/abs/1606.04017}{\texttt{arXiv:1606.04017}}.

\bibitem{Aaboud:2019gxl}
\hrefCMSnoop {}{{ATLAS Collaboration}, ``Measurement of ${\PWpm\PZ}$ production
  cross sections and gauge boson polarisation in ${\Pp\Pp}$ collisions at
  $\sqrt{s}={13\TeV}$ with the {ATLAS} detector'',} \textit{ Eur. Phys. J. C}
  \textbf{ 79} (2019) 535,
  \href{http://dx.doi.org/10.1140/epjc/s10052-019-7027-6}{\doi{10.1140/epjc/s10052-019-7027-6}},
  \href{http://www.arXiv.org/abs/1902.05759}{\texttt{arXiv:1902.05759}}.

\bibitem{CMS:2021pqj}
\hrefCMSnoop {}{{CMS Collaboration}, ``Measurements of the electroweak diboson
  production cross sections in proton-proton collisions at
  $\sqrt{s}={5.02\TeV}$ using leptonic decays'',} \textit{ Phys. Rev. Lett.}
  \textbf{ 127} (2021) 191801,
  \href{http://dx.doi.org/10.1103/PhysRevLett.127.191801}{\doi{10.1103/PhysRevLett.127.191801}},
  \href{http://www.arXiv.org/abs/2107.01137}{\texttt{arXiv:2107.01137}}.

\bibitem{CMS:2021icx}
\hrefCMSnoop {}{{CMS Collaboration}, ``Measurement of the inclusive and
  differential ${\PW\PZ}$ production cross sections, polarization angles, and
  triple gauge couplings in ${\Pp\Pp}$ collisions at $\sqrt{s}={13\TeV}$'',}
  \textit{ JHEP} \textbf{ 07} (2022) 032,
  \href{http://dx.doi.org/10.1007/JHEP07(2022)032}{\doi{10.1007/JHEP07(2022)032}},
  \href{http://www.arXiv.org/abs/2110.11231}{\texttt{arXiv:2110.11231}}.

\bibitem{hepdata}
\hrefCMSnoop {}{}{HEPData} record for this analysis, 2024.
\newblock
  \href{http://dx.doi.org/10.17182/hepdata.155693}{\doi{10.17182/hepdata.155693}}.

\bibitem{Chatrchyan:2008zzk}
\hrefCMSnoop {}{{CMS Collaboration}, ``The {CMS} experiment at the {CERN}
  {LHC}'',} \textit{ JINST} \textbf{ 3} (2008) S08004,
  \href{http://dx.doi.org/10.1088/1748-0221/3/08/S08004}{\doi{10.1088/1748-0221/3/08/S08004}}.

\bibitem{CMS:PRF-21-001}
\hrefCMSnoop {}{{CMS Collaboration}, ``Development of the {CMS} detector for
  the {CERN} {LHC} \mbox{Run 3}'',} \textit{ JINST} \textbf{ 19} (2024) P05064,
  \href{http://dx.doi.org/10.1088/1748-0221/19/05/P05064}{\doi{10.1088/1748-0221/19/05/P05064}},
  \href{http://www.arXiv.org/abs/2309.05466}{\texttt{arXiv:2309.05466}}.

\bibitem{CMS:2020cmk}
\hrefCMSnoop {}{{CMS Collaboration}, ``Performance of the {CMS} {\Lone} trigger
  in proton-proton collisions at $\sqrt{s}={13\TeV}$'',} \textit{ JINST}
  \textbf{ 15} (2020) P10017,
  \href{http://dx.doi.org/10.1088/1748-0221/15/10/P10017}{\doi{10.1088/1748-0221/15/10/P10017}},
  \href{http://www.arXiv.org/abs/2006.10165}{\texttt{arXiv:2006.10165}}.

\bibitem{CMS:2016ngn}
\hrefCMSnoop {}{{CMS Collaboration}, ``The {CMS} trigger system'',} \textit{
  JINST} \textbf{ 12} (2017) P01020,
  \href{http://dx.doi.org/10.1088/1748-0221/12/01/P01020}{\doi{10.1088/1748-0221/12/01/P01020}},
  \href{http://www.arXiv.org/abs/1609.02366}{\texttt{arXiv:1609.02366}}.

\bibitem{CMS:TRG-19-001}
\hrefCMSnoop {}{{CMS Collaboration}, ``Performance of the {CMS} high-level
  trigger during {LHC} \mbox{Run 2}'',} \textit{ JINST} \textbf{ 19} (2024)
  P11021,
  \href{http://dx.doi.org/10.1088/1748-0221/19/11/P11021}{\doi{10.1088/1748-0221/19/11/P11021}},
  \href{http://www.arXiv.org/abs/2410.17038}{\texttt{arXiv:2410.17038}}.

\bibitem{Nason:2004rx}
\hrefCMSnoop {}{P.~Nason, ``A new method for combining {NLO} {QCD} with shower
  {Monte Carlo} algorithms'',} \textit{ JHEP} \textbf{ 11} (2004) 040,
  \href{http://dx.doi.org/10.1088/1126-6708/2004/11/040}{\doi{10.1088/1126-6708/2004/11/040}},
  \href{http://www.arXiv.org/abs/hep-ph/0409146}{\texttt{arXiv:hep-ph/0409146}}.

\bibitem{Frixione:2007vw}
\hrefCMSnoop {}{S.~Frixione, P.~Nason, and C.~Oleari, ``Matching {NLO} {QCD}
  computations with parton shower simulations: the {\POWHEG} method'',}
  \textit{ JHEP} \textbf{ 11} (2007) 070,
  \href{http://dx.doi.org/10.1088/1126-6708/2007/11/070}{\doi{10.1088/1126-6708/2007/11/070}},
  \href{http://www.arXiv.org/abs/0709.2092}{\texttt{arXiv:0709.2092}}.

\bibitem{Alioli:2010xd}
\hrefCMSnoop {}{S.~Alioli, P.~Nason, C.~Oleari, and E.~Re, ``A general
  framework for implementing {NLO} calculations in shower {Monte Carlo}
  programs: the {\POWHEG} \textsc{box}'',} \textit{ JHEP} \textbf{ 06} (2010)
  043,
  \href{http://dx.doi.org/10.1007/JHEP06(2010)043}{\doi{10.1007/JHEP06(2010)043}},
  \href{http://www.arXiv.org/abs/1002.2581}{\texttt{arXiv:1002.2581}}.

\bibitem{Melia:2011tj}
\hrefCMSnoop {}{T.~Melia, P.~Nason, R.~R{\"o}ntsch, and G.~Zanderighi,
  ``${\PWp\PWm}$, ${\PW\PZ}$ and ${\PZ\PZ}$ production in the {\POWHEG}
  \textsc{box}'',} \textit{ JHEP} \textbf{ 11} (2011) 078,
  \href{http://dx.doi.org/10.1007/JHEP11(2011)078}{\doi{10.1007/JHEP11(2011)078}},
  \href{http://www.arXiv.org/abs/1107.5051}{\texttt{arXiv:1107.5051}}.

\bibitem{Nason:2013ydw}
\hrefCMSnoop {}{P.~Nason and G.~Zanderighi, ``${\PWp\PWm}$, ${\PW\PZ}$ and
  ${\PZ\PZ}$ production in the {\POWHEG}-\textsc{box-v2}'',} \textit{ Eur.
  Phys. J. C} \textbf{ 74} (2014) 2702,
  \href{http://dx.doi.org/10.1140/epjc/s10052-013-2702-5}{\doi{10.1140/epjc/s10052-013-2702-5}},
  \href{http://www.arXiv.org/abs/1311.1365}{\texttt{arXiv:1311.1365}}.

\bibitem{Campbell:2010ff}
\hrefCMSnoop {}{J.~M. Campbell and R.~K. Ellis, ``{MCFM for the Tevatron and
  the LHC}'',} \textit{ Nucl. Phys. B Proc. Suppl.} \textbf{ 205-206} (2010)
  10,
  \href{http://dx.doi.org/10.1016/j.nuclphysbps.2010.08.011}{\doi{10.1016/j.nuclphysbps.2010.08.011}},
  \href{http://www.arXiv.org/abs/1007.3492}{\texttt{arXiv:1007.3492}}.

\bibitem{MADGRAPH5}
J.~Alwall\hrefCMSnoop {}{ { et~al.}, ``The automated computation of tree-level
  and next-to-leading order differential cross sections, and their matching to
  parton shower simulations'',} \textit{ JHEP} \textbf{ 07} (2014) 079,
  \href{http://dx.doi.org/10.1007/JHEP07(2014)079}{\doi{10.1007/JHEP07(2014)079}},
  \href{http://www.arXiv.org/abs/1405.0301}{\texttt{arXiv:1405.0301}}.

\bibitem{Ball:2017nwa}
\hrefCMSnoop {}{{NNPDF} Collaboration, ``Parton distributions from
  high-precision collider data'',} \textit{ Eur. Phys. J. C} \textbf{ 77}
  (2017) 663,
  \href{http://dx.doi.org/10.1140/epjc/s10052-017-5199-5}{\doi{10.1140/epjc/s10052-017-5199-5}},
  \href{http://www.arXiv.org/abs/1706.00428}{\texttt{arXiv:1706.00428}}.

\bibitem{Bierlich:2022pfr}
C.~Bierlich\hrefCMSnoop {}{ { et~al.}, ``A comprehensive guide to the physics
  and usage of {\PYTHIA8.3}'',} \textit{ SciPost Phys. Codeb.} \textbf{ 8}
  (2022)
  \href{http://dx.doi.org/10.21468/SciPostPhysCodeb.8}{\doi{10.21468/SciPostPhysCodeb.8}},
  \href{http://www.arXiv.org/abs/2203.11601}{\texttt{arXiv:2203.11601}}.

\bibitem{Sirunyan_2020}
\hrefCMSnoop {}{{CMS Collaboration}, ``Extraction and validation of a new set
  of {CMS} {\PYTHIA8} tunes from underlying-event measurements'',} \textit{
  Eur. Phys. J. C} \textbf{ 80} (2020) 4,
  \href{http://dx.doi.org/10.1140/epjc/s10052-019-7499-4}{\doi{10.1140/epjc/s10052-019-7499-4}},
  \href{http://www.arXiv.org/abs/1903.12179}{\texttt{arXiv:1903.12179}}.

\bibitem{Agostinelli:2002hh}
\hrefCMSnoop {}{{GEANT4} Collaboration, ``{\GEANTfour}---a simulation
  toolkit'',} \textit{ Nucl. Instrum. Meth. A} \textbf{ 506} (2003) 250,
  \href{http://dx.doi.org/10.1016/S0168-9002(03)01368-8}{\doi{10.1016/S0168-9002(03)01368-8}}.

\bibitem{CMS:2017yfk}
\hrefCMSnoop {}{{CMS Collaboration}, ``Particle-flow reconstruction and global
  event description with the {CMS} detector'',} \textit{ JINST} \textbf{ 12}
  (2017) P10003,
  \href{http://dx.doi.org/10.1088/1748-0221/12/10/P10003}{\doi{10.1088/1748-0221/12/10/P10003}},
  \href{http://www.arXiv.org/abs/1706.04965}{\texttt{arXiv:1706.04965}}.

\bibitem{CMS-TDR-15-02}
\hrefCMSnoop {}{{CMS Collaboration}, ``Technical proposal for the {Phase-II}
  upgrade of the {Compact Muon Solenoid}'',} CMS Technical Proposal
  CERN-LHCC-2015-010, CMS-TDR-15-02, 2015.
\newblock
  \href{http://dx.doi.org/10.17181/CERN.VU8I.D59J}{\doi{10.17181/CERN.VU8I.D59J}}.

\bibitem{CMS:2020uim}
\hrefCMSnoop {}{{CMS Collaboration}, ``Electron and photon reconstruction and
  identification with the {CMS} experiment at the {CERN} {LHC}'',} \textit{
  JINST} \textbf{ 16} (2021) P05014,
  \href{http://dx.doi.org/10.1088/1748-0221/16/05/P05014}{\doi{10.1088/1748-0221/16/05/P05014}},
  \href{http://www.arXiv.org/abs/2012.06888}{\texttt{arXiv:2012.06888}}.

\bibitem{CMS:2015xaf}
\hrefCMSnoop {}{{CMS Collaboration}, ``Performance of electron reconstruction
  and selection with the {CMS} detector in proton-proton collisions at
  $\sqrt{s}={8\TeV}$'',} \textit{ JINST} \textbf{ 10} (2015) P06005,
  \href{http://dx.doi.org/10.1088/1748-0221/10/06/P06005}{\doi{10.1088/1748-0221/10/06/P06005}},
  \href{http://www.arXiv.org/abs/1502.02701}{\texttt{arXiv:1502.02701}}.

\bibitem{CMS:2018rym}
\hrefCMSnoop {}{{CMS Collaboration}, ``Performance of the {CMS} muon detector
  and muon reconstruction with proton-proton collisions at
  $\sqrt{s}={13\TeV}$'',} \textit{ JINST} \textbf{ 13} (2018) P06015,
  \href{http://dx.doi.org/10.1088/1748-0221/13/06/P06015}{\doi{10.1088/1748-0221/13/06/P06015}},
  \href{http://www.arXiv.org/abs/1804.04528}{\texttt{arXiv:1804.04528}}.

\bibitem{CMS:2023dsu}
\hrefCMSnoop {}{{CMS Collaboration}, ``Muon identification using multivariate
  techniques in the {CMS} experiment in proton-proton collisions at
  $\sqrt{s}={13\TeV}$'',} \textit{ JINST} \textbf{ 19} (2024) P02031,
  \href{http://dx.doi.org/10.1088/1748-0221/19/02/P02031}{\doi{10.1088/1748-0221/19/02/P02031}},
  \href{http://www.arXiv.org/abs/2310.03844}{\texttt{arXiv:2310.03844}}.

\bibitem{Cacciari:2008gp}
\hrefCMSnoop {}{M.~Cacciari, G.~P. Salam, and G.~Soyez, ``The anti-\kt jet
  clustering algorithm'',} \textit{ JHEP} \textbf{ 04} (2008) 063,
  \href{http://dx.doi.org/10.1088/1126-6708/2008/04/063}{\doi{10.1088/1126-6708/2008/04/063}},
  \href{http://www.arXiv.org/abs/0802.1189}{\texttt{arXiv:0802.1189}}.

\bibitem{Cacciari:2011ma}
\hrefCMSnoop {}{M.~Cacciari, G.~P. Salam, and G.~Soyez, ``{\FASTJET} user
  manual'',} \textit{ Eur. Phys. J. C} \textbf{ 72} (2012) 1896,
  \href{http://dx.doi.org/10.1140/epjc/s10052-012-1896-2}{\doi{10.1140/epjc/s10052-012-1896-2}},
  \href{http://www.arXiv.org/abs/1111.6097}{\texttt{arXiv:1111.6097}}.

\bibitem{Sirunyan:2020foa}
\hrefCMSnoop {}{{CMS Collaboration}, ``Pileup mitigation at {CMS} in {13\TeV}
  data'',} \textit{ JINST} \textbf{ 15} (2020) P09018,
  \href{http://dx.doi.org/10.1088/1748-0221/15/09/P09018}{\doi{10.1088/1748-0221/15/09/P09018}},
  \href{http://www.arXiv.org/abs/2003.00503}{\texttt{arXiv:2003.00503}}.

\bibitem{Bertolini:2014bba}
\hrefCMSnoop {}{D.~Bertolini, P.~Harris, M.~Low, and N.~Tran, ``Pileup per
  particle identification'',} \textit{ JHEP} \textbf{ 10} (2014) 059,
  \href{http://dx.doi.org/10.1007/JHEP10(2014)059}{\doi{10.1007/JHEP10(2014)059}},
  \href{http://www.arXiv.org/abs/1407.6013}{\texttt{arXiv:1407.6013}}.

\bibitem{BTV-16-002}
\hrefCMSnoop {}{{CMS Collaboration}, ``Identification of heavy-flavour jets
  with the {CMS} detector in ${\Pp\Pp}$ collisions at {13\TeV}'',} \textit{
  JINST} \textbf{ 13} (2018) P05011,
  \href{http://dx.doi.org/10.1088/1748-0221/13/05/P05011}{\doi{10.1088/1748-0221/13/05/P05011}},
  \href{http://www.arXiv.org/abs/1712.07158}{\texttt{arXiv:1712.07158}}.

\bibitem{Bols:2020bkb}
E.~Bols\hrefCMSnoop {}{ { et~al.}, ``Jet flavour classification using
  {DeepJet}'',} \textit{ JINST} \textbf{ 15} (2020) P12012,
  \href{http://dx.doi.org/10.1088/1748-0221/15/12/P12012}{\doi{10.1088/1748-0221/15/12/P12012}},
  \href{http://www.arXiv.org/abs/2008.10519}{\texttt{arXiv:2008.10519}}.

\bibitem{CMS:2019ctu}
\hrefCMSnoop {}{{CMS Collaboration}, ``Performance of missing transverse
  momentum reconstruction in proton-proton collisions at $\sqrt{s}={13\TeV}$
  using the {CMS} detector'',} \textit{ JINST} \textbf{ 14} (2019) P07004,
  \href{http://dx.doi.org/10.1088/1748-0221/14/07/P07004}{\doi{10.1088/1748-0221/14/07/P07004}},
  \href{http://www.arXiv.org/abs/1903.06078}{\texttt{arXiv:1903.06078}}.

\bibitem{CMS-DP-2020-018}
\href {https://cds.cern.ch/record/2714938}{{CMS Collaboration}, ``Mitigation of
  anomalous missing transverse momentum measurements in data collected by {CMS}
  at $\sqrt{s}={13\TeV}$ during the {LHC} \mbox{Run 2}'',} Technical Report
  CMS-DP-2020-018, 2020.

\bibitem{ALEPH:2005ab}
\hrefCMSnoop {}{{ALEPH, DELPHI, L3, OPAL, and SLD Collaborations, LEP
  Electroweak Working Group, and SLD Electroweak and Heavy Flavour Groups},
  ``Precision electroweak measurements on the {\PZ} resonance'',} \textit{
  Phys. Rept.} \textbf{ 427} (2006) 257,
  \href{http://dx.doi.org/10.1016/j.physrep.2005.12.006}{\doi{10.1016/j.physrep.2005.12.006}},
  \href{http://www.arXiv.org/abs/hep-ex/0509008}{\texttt{arXiv:hep-ex/0509008}}.

\bibitem{CMS:2018sgc}
\hrefCMSnoop {}{{CMS Collaboration}, ``Observation of single top quark
  production in association with a {\PZ} boson in proton-proton collisions at
  $\sqrt{s}={13\TeV}$'',} \textit{ Phys. Rev. Lett.} \textbf{ 122} (2019)
  132003,
  \href{http://dx.doi.org/10.1103/PhysRevLett.122.132003}{\doi{10.1103/PhysRevLett.122.132003}},
  \href{http://www.arXiv.org/abs/1812.05900}{\texttt{arXiv:1812.05900}}.

\bibitem{CMS:2020hjs}
\hrefCMSnoop {}{{CMS Collaboration}, ``Observation of the production of three
  massive gauge bosons at $\sqrt{s}={13\TeV}$'',} \textit{ Phys. Rev. Lett.}
  \textbf{ 125} (2020) 151802,
  \href{http://dx.doi.org/10.1103/PhysRevLett.125.151802}{\doi{10.1103/PhysRevLett.125.151802}},
  \href{http://www.arXiv.org/abs/2006.11191}{\texttt{arXiv:2006.11191}}.

\bibitem{CMS:2018nsn}
\hrefCMSnoop {}{{CMS Collaboration}, ``Observation of {Higgs} boson decay to
  bottom quarks'',} \textit{ Phys. Rev. Lett.} \textbf{ 121} (2018) 121801,
  \href{http://dx.doi.org/10.1103/PhysRevLett.121.121801}{\doi{10.1103/PhysRevLett.121.121801}},
  \href{http://www.arXiv.org/abs/1808.08242}{\texttt{arXiv:1808.08242}}.

\bibitem{CMS-PAS-EWK-10-002}
\hrefCMSnoop {}{{CMS Collaboration}, ``Measurements of inclusive {\PW} and
  {\PZ} cross sections in ${\Pp\Pp}$ collisions at $\sqrt{s}={7\TeV}$'',}
  \textit{ JHEP} \textbf{ 01} (2011) 080,
  \href{http://dx.doi.org/10.1007/JHEP01(2011)080}{\doi{10.1007/JHEP01(2011)080}},
  \href{http://www.arXiv.org/abs/1012.2466}{\texttt{arXiv:1012.2466}}.

\bibitem{Clopper:1934sws}
\hrefCMSnoop {}{C.~J. Clopper and E.~S. Pearson, ``The use of confidence or
  fiducial limits illustrated in the case of the binomial'',} \textit{
  Biometrika} \textbf{ 26} (1934) 404,
  \href{http://dx.doi.org/10.1093/biomet/26.4.404}{\doi{10.1093/biomet/26.4.404}}.

\bibitem{CMS:2018mlc}
\hrefCMSnoop {}{{CMS Collaboration}, ``{Measurement of the inelastic
  proton-proton cross section at $ \sqrt{s}=13 $ TeV}'',} \textit{ JHEP}
  \textbf{ 07} (2018) 161,
  \href{http://dx.doi.org/10.1007/JHEP07(2018)161}{\doi{10.1007/JHEP07(2018)161}},
  \href{http://www.arXiv.org/abs/1802.02613}{\texttt{arXiv:1802.02613}}.

\bibitem{CMS-PAS-LUM-22-001}
\href {https://cds.cern.ch/record/2890833}{{CMS Collaboration}, ``Luminosity
  measurement in proton-proton collisions at {13.6\TeV} in 2022 at {CMS}'',}
  CMS Physics Analysis Summary CMS-PAS-LUM-22-001, 2024.

\bibitem{CMS:LUM-17-003}
\hrefCMSnoop {}{{CMS Collaboration}, ``Precision luminosity measurement in
  proton-proton collisions at $\sqrt{s}={13\TeV}$ in 2015 and 2016 at {CMS}'',}
  \textit{ Eur. Phys. J. C} \textbf{ 81} (2021) 800,
  \href{http://dx.doi.org/10.1140/epjc/s10052-021-09538-2}{\doi{10.1140/epjc/s10052-021-09538-2}},
  \href{http://www.arXiv.org/abs/2104.01927}{\texttt{arXiv:2104.01927}}.

\bibitem{Barlow:1993dm}
\hrefCMSnoop {}{R.~Barlow and C.~Beeston, ``Fitting using finite {Monte Carlo}
  samples'',} \textit{ Comput. Phys. Commun.} \textbf{ 77} (1993) 219,
  \href{http://dx.doi.org/10.1016/0010-4655(93)90005-W}{\doi{10.1016/0010-4655(93)90005-W}}.

\bibitem{CMS:2024onh}
\hrefCMSnoop {}{{CMS Collaboration}, ``The {CMS} statistical analysis and
  combination tool: \textsc{combine}'',} \textit{ Comput. Softw. Big Sci.}
  \textbf{ 8} (2024) 19,
  \href{http://dx.doi.org/10.1007/s41781-024-00121-4}{\doi{10.1007/s41781-024-00121-4}},
  \href{http://www.arXiv.org/abs/2404.06614}{\texttt{arXiv:2404.06614}}.

\bibitem{Grazzini:2019jkl}
M.~Grazzini\hrefCMSnoop {}{ { et~al.}, ``{NNLO} {QCD} + {NLO} {EW} with
  \textsc{matrix}+\textsc{OpenLoops}: precise predictions for vector-boson pair
  production'',} \textit{ JHEP} \textbf{ 02} (2020) 087,
  \href{http://dx.doi.org/10.1007/JHEP02(2020)087}{\doi{10.1007/JHEP02(2020)087}},
  \href{http://www.arXiv.org/abs/1912.00068}{\texttt{arXiv:1912.00068}}.

\end{thebibliography}\endgroup
\cleardoublepage \appendix\section{The CMS Collaboration \label{app:collab}}\begin{sloppypar}\hyphenpenalty=5000\widowpenalty=500\clubpenalty=5000
\cmsinstitute{Yerevan Physics Institute, Yerevan, Armenia}
{\tolerance=6000
A.~Hayrapetyan, A.~Tumasyan\cmsAuthorMark{1}\cmsorcid{0009-0000-0684-6742}
\par}
\cmsinstitute{Institut f\"{u}r Hochenergiephysik, Vienna, Austria}
{\tolerance=6000
W.~Adam\cmsorcid{0000-0001-9099-4341}, J.W.~Andrejkovic, L.~Benato\cmsorcid{0000-0001-5135-7489}, T.~Bergauer\cmsorcid{0000-0002-5786-0293}, S.~Chatterjee\cmsorcid{0000-0003-2660-0349}, K.~Damanakis\cmsorcid{0000-0001-5389-2872}, M.~Dragicevic\cmsorcid{0000-0003-1967-6783}, P.S.~Hussain\cmsorcid{0000-0002-4825-5278}, M.~Jeitler\cmsAuthorMark{2}\cmsorcid{0000-0002-5141-9560}, N.~Krammer\cmsorcid{0000-0002-0548-0985}, A.~Li\cmsorcid{0000-0002-4547-116X}, D.~Liko\cmsorcid{0000-0002-3380-473X}, I.~Mikulec\cmsorcid{0000-0003-0385-2746}, J.~Schieck\cmsAuthorMark{2}\cmsorcid{0000-0002-1058-8093}, R.~Sch\"{o}fbeck\cmsAuthorMark{2}\cmsorcid{0000-0002-2332-8784}, D.~Schwarz\cmsorcid{0000-0002-3821-7331}, M.~Sonawane\cmsorcid{0000-0003-0510-7010}, W.~Waltenberger\cmsorcid{0000-0002-6215-7228}, C.-E.~Wulz\cmsAuthorMark{2}\cmsorcid{0000-0001-9226-5812}
% ...
\par}
\vskip\cmsinstskip
\dag:~Deceased\\
$^{1}$Also at Yerevan State University, Yerevan, Armenia\\
$^{2}$Also at TU Wien, Vienna, Austria\\
% ...
\end{sloppypar}
\end{document}